\documentclass[revtex4, numberedappendix]{openjournal}

\usepackage{graphicx}
\usepackage{dcolumn}
\usepackage{bm}

\usepackage{hyperref}
\usepackage{amsmath}
\usepackage{braket}
\usepackage{xcolor}

\newcommand{\create}[1]{\hat{#1}^{\dagger}}

\graphicspath{{./}{figures/}}

\shorttitle{Two-Photon Precision Astrometry}
\shortauthors{Stankus et al.}

\begin{document}

%
\title{Two-photon amplitude interferometry for 
precision astrometry}


\author{Paul Stankus}
\author{Andrei Nomerotski}
\author{An\v{z}e Slosar}
\affiliation{Brookhaven National Laboratory, Upton NY 11973, USA}

\author{Stephen Vintskevich}
\affiliation{Moscow Institute of Physics and Technology,
Dolgoprudny, Moscow Region 141700, Russia}

%
%

%
\begin{abstract}
Improved quantum sensing of photons from astronomical objects 
could provide high resolution observations in the optical benefiting numerous fields, including general relativity, dark matter studies, and cosmology.  It has been recently proposed that stations in optical interferometers would not require a phase-stable optical link if instead sources of quantum-mechanically entangled pairs could be provided to them, potentially enabling hitherto prohibitively long baselines.  A new refinement of this idea is developed, in which two photons from different sources are interfered at two separate and  decoupled stations, requiring only a slow classical information link between them. We rigorously calculate the observables and contrast this new interferometric technique with the Hanbury Brown \& Twiss intensity interferometry.  We argue this technique could allow robust high-precision measurements of the relative astrometry of the two sources.  A basic calculation suggests that angular precision on the order of $10$~microarcsecond in the relative opening angle could be achieved in a single night's observation of two bright stars.
\end{abstract}

%


%
\section{Introduction}
\label{sec:introduction}

Quantum phenomena are often strange and non-intuitive effects that happen only in the atomic world. At the core of them is entanglement, which has no counterparts in our classical world, and which is enabling new measurement techniques and devices beyond what can be achieved classically. The next technological frontiers will exploit these quantum phenomena to augment sensitivity and to overcome fundamental limitations in macroscopic systems. Harnessing quantum effects has already proven to be groundbreaking in many experiments: to name one dramatic example, in the Laser Interferometer Gravitational-Wave Observatory (LIGO) photon shot noise imposes a fundamental limit on the sensitivity of the km-length interferometers; but it was possible to exploit quantum interactions to overcome this limit by using the higher sensitivity of squeezed states of light \cite{Tse2019, Buonanno2001}.

Traditional Michelson optical interferometers are essentially classical, in that they can -- and were, of course -- described in terms a Maxwellian electro-magnetic (EM) wave interfering with itself after taking two paths; this translates perfectly to a quantum description of one photon interfering with itself.  However, there is a new wave of interest in interferometry using multiple photons, whose mechanisms are inherently quantum mechanical, which offer the prospects of increased baselines and finer resolutions.  We will discuss recent ideas for quantum-assisted interferometry using the resource of entangled pairs, and specifically a two-photon amplitude technique aimed at improved precision in dynamic astrometry. 

It was pointed out by Gottesman, Jennewein and Croke \cite{Gottesman2012} in 2012 that optical interferometer baselines could be extended, without an optical connecting path, if a supply of entangled photon states between the two stations could be provided. If these states could then be interfered locally at each station with an astronomical photon that has impinged on both stations, the outcomes at the two stations would be correlated in a way that is sensitive to the phase difference in the two paths of the photon, thus reproducing the action of an interferometer. Equivalently, this can be seen as using an interference measurement at one station to teleport the state of that station’s astronomical photon to the other station, and interfering it with its counterpart there.
This teleportation technique would allow to uncouple the two observing stations, in principle then allowing arbitrary baselines and much finer angular resolution scales, down to the micro-arcsecond level or below. 


In this work we extended the above idea to use the second photon produced by another astronomical sky source. The path length difference between the two photons leads to a phase offset and if the two photons are close enough together in both time and frequency, then due to quantum mechanical interference the pattern of coincidences in the two stations will be sensitive to the phase differences, and this in turn will be sensitive to the relative opening angle between the two sources. In this scheme no optical connection path is needed between the two stations, a major simplification of the original idea; and the measurement can be carried out in many spectroscopic bins simultaneously. 


The relation between the observed pair rate correlations and the sources' sky positions and extent is calculated using quantum optics methods, and shows that the two-source technique can be seen as a generalization of the traditional Hanbury Brown \& Twiss~(HBT) intensity interferometry from a single source.  Our new technique produces alternating correlations and anti-correlations between measurement outcomes from different receivers, e.g. fringes, and so we term it two-photon amplitude interferometry.  We then propose an Earth-rotation scan observation program and show that the source opening angle can be derived directly from a measurement of the fringe passing rate in the detected pair correlations.  Day-over-day comparisons will then measure pair relative motions stemming from parallax, proper motions, orbital motions, gravitational lensing, etc.  We derive the statistical sensitivity of the technique with a Fisher information matrix approach, and calculate that a nominal experiment could reach a precision on the order of 10~$\mu$as on the opening angle between two bright stars in a single night's observation.


\medskip
The text below is organized as follows: in Section~\ref{sec:astro_motivation} we detail how improvements in astrometric measurements can impact science topics in cosmology and astrophysics. In 
Section~\ref{sec:astrometry_interferometry} 
we recap limitations of classical interferometry in astronomical context.
In Section~\ref{sec:basics_two_photon_amplitude} we introduce the new quantum technique of two-photon amplitude interferometry and its application for high-resolution astrometry.  In Section~\ref{sec:field_theory_calc} we provide rigorous derivations of the technique employing calculations within quantum optics theory.  Finally in Section~\ref{sec:sky_observables} we propose new observables for practical implementation of the technique and evaluate its precision for a bright star example, and discuss scaling relations relevant for dim objects.  We accompany the main text with supporting detailed calculations for the above sections in Appendices~\ref{app:classical}, \ref{app:field} and~\ref{app:sensitivity_estimate}.

%
\section{Motivation for precision relative astrometry}
\label{sec:astro_motivation}

It is impossible to foresee all the scientific opportunities offered by greatly improved astrometric resolution compared to current techniques.  Here we consider a few example cases.

\smallskip

\textbf{Precision parallax and cosmic distance ladder}: There is presently a tension in determination of the present-day expansion rate of the Universe, also known as the Hubble parameter $H_{0}$, between those based on distance ladder in the local Universe and those based on indirect extrapolation from higher redshift measurements of Baryonic Acoustic Oscillations and Cosmic Microwave Background. The distance ladder method uses a set of probes to bootstrap distance calibration from local measurements to cosmological distances. Parallaxes are used to calibrate distance to the Cepheid variable stars, which have a fixed period-luminosity relation. Cepheid calibration is then transferred from our own galaxy to other galaxies where supernovae Type Ia are observed, and supernovae Type Ia in somewhat more distant galaxies are then used for $H_{0}$ measurement~\cite{Ash1967, Linden2009}.

Naturally, however, errors in any one step affect the entire ladder. Direct parallax measurements are systematically very robust, but are necessarily limited by the achievable astrometric precision. The most sensitive astrometric data with precision of few dozen microarcsec is provided by the recent GAIA space mission~\cite{Katz2019, Lindegren2021, Groenewegen2021}.  The use of Cepheids as standard candles in the distance ladder is complicated by a number of systematic uncertainties in their period-luminosity dependence. An improvement in the astrometric precision, possibly by several orders of magnitude, as proposed here could allow us to completely sidestep the Cepheids and use parallax directly on galaxies with supernovae Type Ia, providing a landmark advance in $H_{0}$ measurements. In practice this will be done by measuring the fringe changes from a pair of objects composed of a “background object” such as a distant quasar that is essentially fixed on the celestical sphere and a “foreground” object that is subject to parallactic correction as the Earth orbits the sun.

\smallskip

\textbf{Precision binary orbits}: Measurement of the orbits of binary systems could provide a method of determining distances to objects within the Galaxy, completely independent of parallax.  Broadly speaking, spectroscopy can be used to measure the velocities of one or both objects in their orbits, and also the period of a binary's orbits, and these can then be combined to determine the absolute size of the orbit.  A precision astrometric measurement of one or both objects' apparent orbital motion can then be combined with the orbital size to infer the distance to the binary.  Such a technique could provide an independent calibration of the first rung of the distance ladder, with a very different set of systematics than from the parallax method.

This technique has been used to estimate distances to binaries in cases where both objects can be imaged in the same photon mode, in the context of both Michelson~\cite{hummel1998navy}  and Hanbury Brown Twiss interferometry~\cite{HerbisonEvans1971}.  The technique described in this paper will extend the possibility to systems where only one object is bright, since its motion can be measured against any convenient bright reference object well outside the same mode.

\smallskip
\textbf{Mapping microlensing events}: The nature of dark matter (DM) remains one of the greatest mysteries of the Universe. One possibility is that DM exists in the form of compact objects the size of planets or stars, perhaps as black holes, or just extended virialized subhalos of dark matter particles. Such objects act as gravitational microlenses both in the Galaxy and in extragalactic lens systems. Traditionally, microlensing has been observed photometrically by looking at the apparent change in brightness of object during passage of the lens in front of it. However, the main signature would be measurement of the change in position and shape of the object, which has so far eluded astrometric measurements due to lack of precision \cite{Erickcek2011, Wyrzykowski2016}.  With lensing a star’s image is split into two loci, each moving and  evolving while the star moves in the plane behind the lens \cite{Sahu2014}.
Improving the astrometric precision of the measurements will allow to decrease the detection thresholds, potentially increasing the statistics hence the sensitivity to the DM subhalos. The astrometric approach is also more straightforward to interpret in terms of the lens mass and its spatial distribution. An interesting novel possibility here would be to constrain astrometric jitter that would in turn constrain the presence of a population of small microlenses in a statistical manner.

\smallskip
\textbf{Peculiar motions and dark matter}: It is well known that the dynamics of our Galaxy are affected by the DM distribution in the Galaxy. The redshifts and blueshifts of stars measure their radial velocities and are technically feasible for all bright stars across the Galaxy. The transverse velocities, on the other hand, are probed by measuring peculiar motions of stars through astrometric measurements and are currently available only in the vicinity of the Earth \cite{Steinmetz2020, Simon2018, Katz2019}. Thus the reconstruction of the truly 3D velocities for a substantial sample of stars in the Galaxy is not possible now. Measurement of the full 3D velocity vector for a significant portion of the stars across the Galaxy would allow us to infer the gravitational potential for the galactic halo and would be transformative. It will give us a census of merging events in the history of the milky way halo and it would directly probe dark matter self-interaction, its interactions with baryons and other exciting possibilities (see e.g. \cite{Chu2019}). It may also allow to detect populations of dark matter subhalos, as well as density fluctuations sourced by the dark matter \cite{Tilburg2018, MishraSharma2020, Gardner2021}, and to measure clumping on a range of scales not available through other means, giving direct constraints on the coldness of dark matter \cite{Majewski2007}.

\smallskip
\textbf{Further}: Much improved astrometric precision also will offer large gains in other areas of astrophysics, which are very important for modern science. For example, it could revolutionize searches for exoplanets and interpretation of their properties through direct observation of disturbed trajectories for their host stars, or even by directly resolving the star-planet binary systems. Precision astrometry can also be used for the detection of gravitational waves as coherent movements of stars.  Many more applications can also be imagined.

%
\section{Classical single-photon interferometry}
\label{sec:astrometry_interferometry}

%
\begin{figure}
\begin{center}
\includegraphics[width=0.6\linewidth]{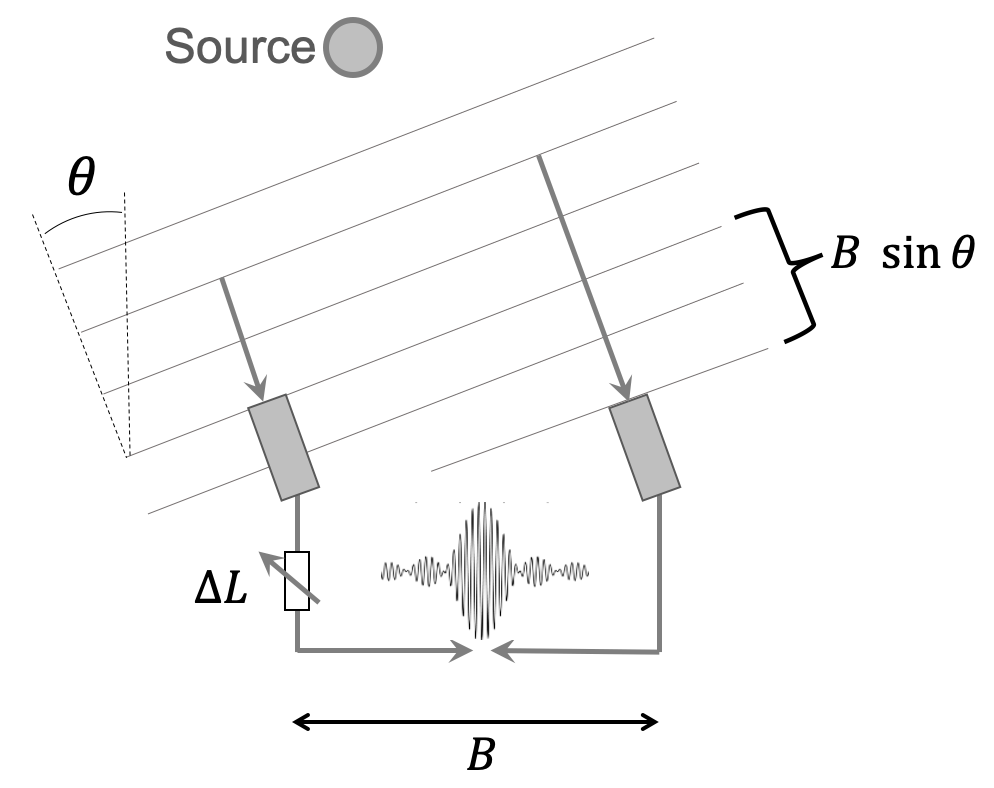}
\caption{Traditional (Michelson) stellar interferometry. A single photon from an astronomical source impinges on two telescopes nearly simultaneously, with a phase difference
determined by the difference in path lengths.
The two optical paths are brought together across the
baseline, where the photon's interference with itself
depends on the path length difference and hence on
the direction to the source.  Inteferometry is 
generally sensitive to structures with angular
scales on the order of $\Delta \theta \sim \lambda/B$
where $B$ is the baseline length and $\lambda$ is
the photon wavelength. 
}
\label{fig:interferometer_basic}
\end{center}
\end{figure}

The basic figure of merit for any astrometry instrument will be the scale of its angular resolution, which will determine the smallest feature size, or change in  feature position, that it can observe.  The resolution of a single aperture is diffraction-limited at a scale of $\Delta\theta \sim \lambda/D$ where $\lambda$ is the photon wavelength and $D$ is the aperture width.  Interferometers
can access finer resolution by using separate sub-apertures across baselines larger than any single aperture, and interferometry is a well recognized tool for precision astrometry~\cite{tenBrummelaar2005, Pedretti2009, Robbe-Dubois2014}.

Figure~\ref{fig:interferometer_basic} illustrates
the prototypical two-element optical inteferometer, which was pioneered by Michelson and Fizeau starting in the 1890's and first used to measure stellar diameters in 1920 \cite{Michelson1887, Michelson1921}.  A single photon is focused, separately, into each of two entrances and the two optical paths are then brought together across the baseline.  At this point the photon wave interferes with itself, and the result over many photons is a fringe pattern that is sensitive to source intensity variations on the scale of $\Delta\theta \sim \lambda/B$, where $B$
is the separation between the sub-apertures.  Even though single quanta are involved the operation of a Michelson stellar interferometer is essentially classical and can be completely described in terms of Maxwellian EM waves.

Single-photon optical interferometry is completely analogous with radio interferometry. In radio $\lambda$ can be on the order of meters to millimeters with baselines $B$ of thousands
of kilometers in VLBI (Very Long Baseline Interferometry) with observatories spread across  the Earth.  Astronomical VLBI is greatly enabled by the fact that radio-frequency EM waveforms can be recorded independently and interfered offline later, which is not the case in the optical.  Radio VLBI has provided some of the most high-resolution observations in astronomy, dramatically including the recent imaging of a supermassive black hole in M87 \cite{2019ApJ...875L...2E}. 

While very successful in radio frequency domain, long baseline interferometry has not been developed in the optical domain to nearly the same scales.  The longest baselines in optical interferometers are in the range of a few~$\times 10^2$~meters (130~m at the VLTI and 330~m CHARA, for example) \cite{2012A&ARv..20...53B}. 
This is partly due simply to the choice of which feature size is interesting for a particular astrophysics goal; for a 1~$\mu$m wavelength across a 100~m baseline $\lambda/B \sim 10^{-8} \sim$~milli-arcsecond~\cite{Robbe-Dubois2014} 
\cite{tenBrummelaar2005, Pedretti2009, Martinod2018, Gillessen2009MONITORINGSO, hees2017testing, do2019relativistic}.
At the same time, though, Michelson interferometry requires maintaining an optical path which is stable at the sub-wavelength level; and extending such a pathway, typically a vacuum tunnel, to allow baselines of $10^{3}$--$10^{5}$~m or more will involve substantial costs.  As we discuss below, however, one can exploit quantum effects to uncouple the two observing stations, in principle then allowing arbitrary baselines and much finer angular resolution scales, down to the micro-arcsecond level or below.

%
\section{Two-photon amplitude interferometry}
\label{sec:basics_two_photon_amplitude}

In classical single-photon interferometry we are observing a single source photon in multiple stations. The main focus of this paper is a two-photon interferometry, a novel technique where two photons from two sources are interfered with sensitivity to their relative phases.

We will shift to using a quantum description of interferometry in two stages.  In this Section we will quickly lay out the basics of two-photon, two-source amplitude interferometry using a simple quantum mechanical picture of monochromatic photons as particles, e.g. definite Fock states, carried forward in a Shr\"{o}dinger representation.  
Then in Section~\ref{sec:field_theory_calc} we will go through a full calculation with time-dependent electric field operators.  This allows us to address properly the quasi-monochromatic case and the time correlations between the measurements of the two photons, as well as extended sources.

While the two-photon optical techniques relies on purely quantum effects for detection there are classical analogues, much like in the case of single photon amplitude interferometry. We discuss properties of these two-source classical interferometers and relation to the standard interferometry in the Appendix~\ref{app:classical}.

%
\subsection{Single-source amplitude interferometry}
\label{subsec:single_source_interf}

\begin{figure}
\begin{center}
\includegraphics[width=0.60\linewidth]{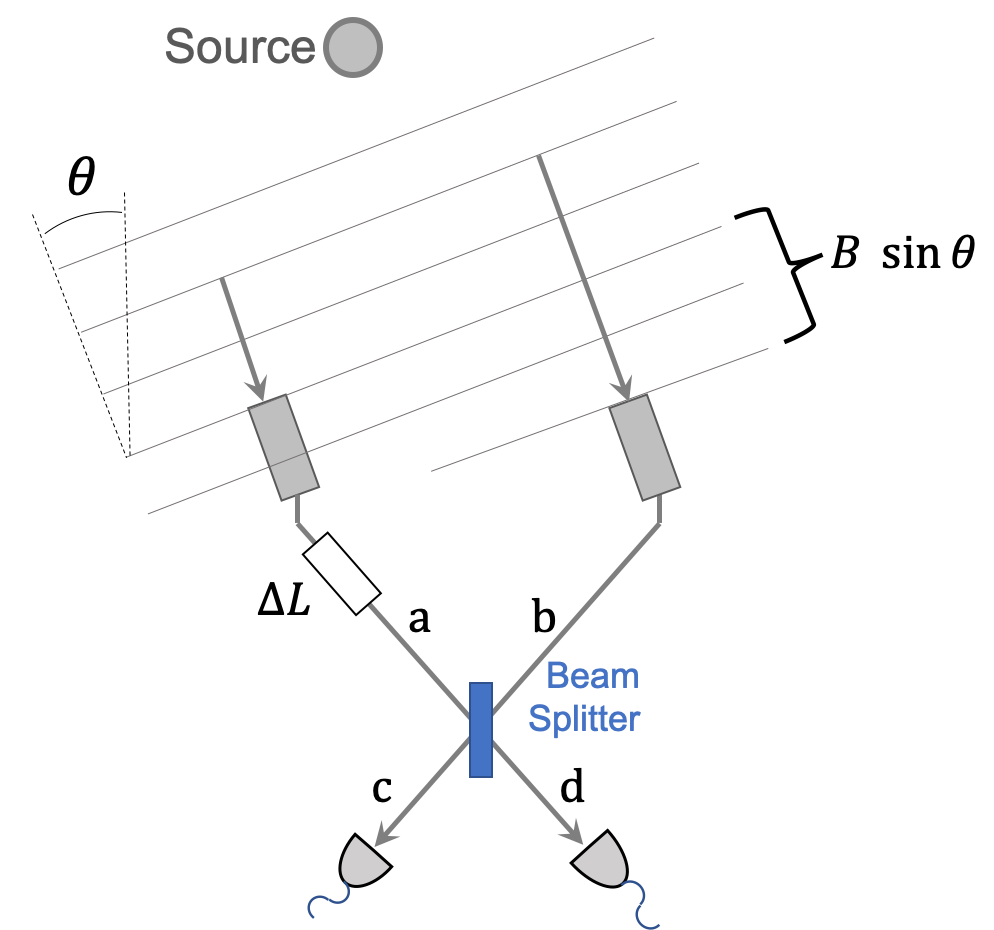}
\caption{Single-photon description of a beam splitter interferometer.  The photon comes down from a point source and enters both telescopes as a plane wave; $B$ is the baseline distance between the stations and $\theta$ is the equatorial polar angle of the source relative to the axis of the baseline.  The photon's entrance leads to a superposition of exciting the input modes ``a'' and ``b'' of a symmetric beam splitter, whose output modes ``c'' and ``d'' are each viewed by detectors.  The effective path length difference from the wavefront to the splitter is the sum of the free-space difference $B \sin \theta$ and some $\Delta L$ in the combining apparatus.}
\label{fig:single_photon_beamsplitter}
\end{center}
\end{figure}

Following Gottesman, Jennewein and Croke (GJC) \cite{Gottesman2012},
in developing our quantum description it is useful to first revisit the traditional Michelson stellar interferometer, re-drawn in Figure~\ref{fig:single_photon_beamsplitter} as a beam splitter interferometer (BSI).  Here we imagine a simplified situation where a single, monochromatic photon from a point source comes down as a plane wave, i.e. in a pure Fock state.  This then impinges on both receiving stations where it excites a superposition of the single modes $a$ and $b$, which are the input channels of a symmetric beam splitter.  Assuming both receiving telescope systems are identical\footnote{For simplicity we are neglecting questions of polarization, and simply assume that the combining apparatus steers the common polarization of the plane wave into common polarizations at the splitter.}, then up to an overall phase the photon state at the entrance to the splitter will be:
\begin{eqnarray}
    \ket{\Psi_\mathrm{Init}} & = & 
    \frac{1}{\sqrt{2}} \left( \ket{1_{a} 0_{b}} + e^{i \delta}\ket{0_{a} 1_{b}} \right)
 = \frac{1}{\sqrt{2}} (\create{a} + e^{i \delta}\create{b}) \ket{vac} \, \equiv \create{\sigma} \ket{vac}
\label{eq:initial_single_sky_photon}
\end{eqnarray}

\noindent
where $\create{a}$ and $\create{b}$ are creation operators for their respective modes, with similar notation for other labelled modes;  $\ket{vac}$ is the state with no excitations; and the phase difference $\delta$ is determined by the effective difference in path lengths from the wavefront to the splitter. Since the modes $a$ and $b$ have identical frequencies we can then define the new mode $\sigma$ which is excited by the sky photon, with its creation operator being $\create{\sigma} \equiv (\create{a} + e^{i \delta}\create{b})/\sqrt{2}$.

%

As in Figure~\ref{fig:single_photon_beamsplitter} the two modes $a$ and $b$ are directed as inputs to a symmetric beam splitter with output channels $c$ and $d$.  With a convenient choice of phase convention we can write the evolution of the state through the beam splitter simply as the substitutions
\begin{equation}
\create{a} \rightarrow  (\create{c} + \create{d})/\sqrt{2} \qquad
\create{b} \rightarrow (\create{c} - \create{d})/\sqrt{2}
\label{eq:single_beam_splitter_transforms} 
\end{equation}
\noindent
which leads to final state
\begin{equation}
\ket{\Psi_\mathrm{Final}} =  \frac{1}{2}
    \left(  
       (1+e^{i \delta}) \create{c} +  (1-e^{i \delta})\create{d} 
     \right)
    \ket{vac}
\label{eq:single_photon_BS_final_state}
\end{equation}

\noindent
from which we can read off the probabilities of having the photon measured in the ``c'' detector versus the ``d'' detector:
\begin{eqnarray}
   P(c) & =& 
      \frac{1}{4} |1+e^{i \delta}|^2 = \frac{1}{2} 
        \left( 1+\cos(\delta) \right) 
      \nonumber \\
   P(d) & =& 
      \frac{1}{4} |1-e^{i \delta}|^2 = \frac{1}{2}
         \left( 1-\cos(\delta) \right) 
\label{eq:prob_one_photon}
\end{eqnarray}

\noindent
This is exactly the result one would expect, once we note that the BSI is optically equivalent to a Mach-Zehnder interferometer and so sinusoidally sensitive to the phase difference along the two paths.

Following Figure~\ref{fig:single_photon_beamsplitter} we can write $\delta$ in terms of the path length difference, i.e. $\delta = 2 \pi (B \sin\theta - \Delta L)/\lambda$, where $B$ is the baseline length, $\theta$ marks the position of the point source in the sky, $\lambda$ is the photon wavelength, and $\Delta L$ is an effective path length difference within the combining apparatus.
We can define a useful sky observable $O_\mathrm{BSI}$ for the beam splitter interferometer from the numbers $N(c)$ of photons detected at $c$ versus $N(d)$ at $d$ and its expectation value over some short period of time:
\begin{equation*}
    O_\mathrm{BSI}  \equiv  \frac{N(c)-N(d)}{N(c)+N(d)}
\end{equation*}
\noindent
In the case that the source is small and the time interval of the observation is short $\theta$ is essentially single-valued; and if $\Delta L$ is stable to better than a wavelength over that period then $O_\mathrm{BSI}$ will average to
\begin{equation}
    \langle O_\mathrm{BSI} \rangle = 
    \cos \left( \frac{2 \pi B \sin\theta}{\lambda} 
    - \frac{2 \pi \Delta L}{\lambda} \right)
\label{eq:Observable_BSI}
\end{equation}
\noindent
Equation~\eqref{eq:Observable_BSI} shows the extreme sensitivity of the observable to the source's sky position, amplified by a factor on the order of $B/\lambda$, typically in the range $10^6$ -- $10^9$ for optical interferometry.

Further, if we construct the $O_\mathrm{BSI}$ from observing an extended source the result will be sensitive to a Fourier moment of the source's distribution across the sky at a wavenumber of $2 \pi B\cos\theta/\lambda$.  If $\Delta L$ is varied quickly in a controlled manner, or $\cos \theta$ is allowed to vary, then the fringes in $O_\mathrm{BSI}$ will trace out the amplitude and phase of the Fourier component.

%
\subsection{Double-source amplitude interferometry}
\label{subsec:double_source_interf}

With the review of single-photon interferometry from a single source in beam splitter form we can now readily describe the new technique of two-photon interferometry from two sky sources.

\begin{figure}
\begin{center}
\includegraphics[width=0.60\linewidth]{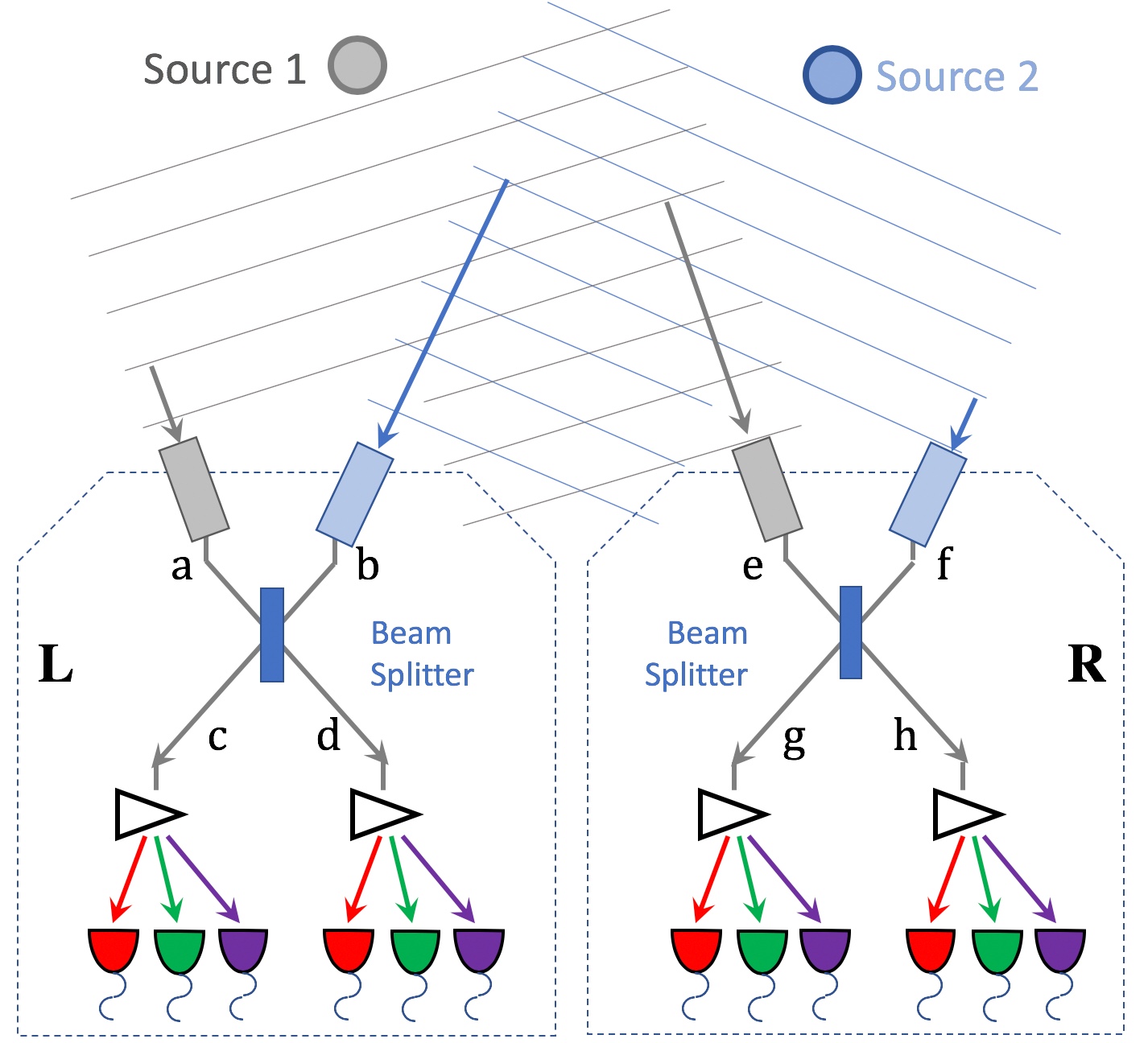}
\caption{The two-photon, two source amplitude interferometer. Source 1 sends a photon which arrives as a plane wave at both input single-mode channels ``a'' and ``e''.  The path length difference leads to a phase offset of $\delta_{1}$, and the photon is in an entangled state 
$| 0 \rangle_{L} | 1 \rangle_{R} + e^{i \delta_{1}}  \, | 1 \rangle_{L} | 0 \rangle_{R}$ between the two observatories \textbf{L} and \textbf{R}
(we recommend references~\cite{Gulbahar2020,Tan1991,Hardy1994,vanEnk2005,Morin2013,Lvovsky2001} for details of the mode and path entanglement phenomena of photons).
At the same time a photon from Source 2 enters channels ``b'' and  ``f'' with a phase difference $\delta_2$ and in an entangled state $| 0 \rangle_{L} | 1 \rangle_{R} + e^{i \delta_{2}}  \, | 1 \rangle_{L} | 0 \rangle_{R}$.  (The photon collection at each station can be in two separate telescopes, as shown, or in one, as long as the two sources can be imaged separately.)  These are then interfered using the beam splitters in the two stations as shown.  If the two photons are close enough together in both time and frequency, then due to quantum mechanical interference the pattern of coincidences between measurements at ``c'' and ``d'' in \textbf{L} and ``g'' and ``h'' in \textbf{R} will be sensitive to the {\em difference} in phase differences $(\delta_{1} - \delta_{2})$; and this in turn will be sensitive to the relative opening angle between the two sources.  No optical connection path is needed between the two stations; and the measurement can be carried out in many spectroscopic bins simultaneously, as suggested by the arrays of detectors at each output.}
\label{fig:interferometer_spread_detectors_wavefront}
\end{center}
\end{figure}

Figure~\ref{fig:interferometer_spread_detectors_wavefront}
shows the basic arrangement: the two sources~1 and~2 are both observed from each of two stations, \textbf{L} and~\textbf{R}.  For sources widely separated on the sky we can imagine this being done with four telescopes, as shown; while for sources very close together they could both be imaged in the same telescope system at one station, as long as they can be separated in the focal plane.  The key requirement is that photons from Source~1 be channeled into single spatial modes~$a$ at station~\textbf{L} and~$e$ at station~\textbf{R}; while those from Source~2 are separately collected into the two single spatial modes $b$ and $f$ as shown.

The photon modes~$a$ and~$b$ at station~\textbf{L} are then brought to the inputs of a symmetric beam splitter, with output modes labelled $c$ and $d$; and the same for input modes $e$ and $f$ split onto output modes $g$ and $h$ at station~\textbf{R}.  The four outputs are then each viewed by a fast, single-photon sensitive detector.  We imagine that the light in each output port is spectrographically divided into very small bins -- we imagine using bandwidths on the order of $\Delta\nu\sim$1~GHz wide, see Section~\ref{subsec:bright_star_example} below -- and each spectral bin then constitutes a separate experiment with four detectors.  Note, for convenience we are ignoring polarization degrees of freedom, effectively assuming that only one mode in each channel is being used.

The basic observational event will be the registration of two photons in the system close enough together in time and frequency to be in the same temporal mode, and so be indistinguishable (see Sections~\ref{subsec:field_theory_two-photon} and~\ref{subsec:bright_star_example} below).  Then the pattern of coincidences between detectors at the two stations, which can be compared after the fact via some classical network transport, will reveal information about the sources' positions and extent.

We first decide to post-select final states with
exactly two photons present across the output channels
$c,d,g,h$.  Then in the particle description we need
consider three types of initial two-photon state,
namely that with one photon from each source and those
with two photons from the same source.  In each case
the wavefunction at the entrance to the splitters can
be written as simple Fock states:
\begin{eqnarray}
\text{Two from source 1: } & &
\ket{\Psi_\mathrm{Init}^{11}} 
    =  \frac{1}{\sqrt{2}}
    \create{\sigma_1} \create{\sigma_1} \ket{vac}  \nonumber \\
\text{Two from source 2: } & &
\ket{\Psi_\mathrm{Init}^{22}} 
    = \frac{1}{\sqrt{2}}
    \create{\sigma_2} \create{\sigma_2} \ket{vac}  \nonumber \\
\text{One from each source: } & &
\ket{\Psi_\mathrm{Init}^{12}} 
    =  \create{\sigma_1} \create{\sigma_2} \ket{vac} \label{eq:initial_two_sky_photons} 
\end{eqnarray}
\noindent
where $\create{\sigma_1}$ and $\create{\sigma_2}$ are the creation operators from the two point sources into the input channels of two beam splitters.  Generalizing directly from  Equation~\eqref{eq:initial_single_sky_photon}, up to an arbitrary overall phase for each operator we have
\begin{equation}
\create{\sigma_1}  \equiv  (\create{a} 
+ e^{i \delta_{1}} \create{e})/ \sqrt{2} \,\,\,\,\,\,
\create{\sigma_2}  \equiv  (\create{b} 
+ e^{i \delta_{2}} \create{f})/ \sqrt{2} 
\label{eq:sigma_1_sigma_2_def}
\end{equation}

We can write the final states which follow from the initial states listed in Equation~\eqref{eq:initial_two_sky_photons} as above, by expanding the $\create{\sigma}$'s as in Equation~\eqref{eq:sigma_1_sigma_2_def} and then modelling the beam splitters' actions as in Equation~\eqref{eq:single_beam_splitter_transforms} for the $a$ and $b$ channels and the corresponding versions for the $e$ and $f$ channels.  The resulting final state for the case of two photons from Source~1 is:
\begin{eqnarray}
  \ket{\Psi_\mathrm{Final}^{11}}  = \frac{1}{2 \sqrt{2}} (
     \frac{1}{2} ( \create{c} \create{c} +
    \create{d} \create{d} +
    e^{2 i \delta_1} (  \create{g} \create{g} +
    \create{h} \create{h})) +
    \create{c} \create{d} + e^{i \delta_1} (  \create{c} \create{g} +
    \create{c} \create{h} + 
     \create{d} \create{g} +
    \create{d} \create{h}) +  e^{2 i \delta_1} \create{g} \create{h}
    ) \ket{vac}
\end{eqnarray}

\noindent
From this we can read off the probabilities for each of the two-photon outcomes:
\begin{eqnarray}
&P_{11}(cc)=P_{11}(dd)=P_{11}(gg)=P_{11}(hh)  =  1/16 \nonumber \\
&P_{11}(cd)=P_{11}(cg)=P_{11}(ch)  = 
P_{11}(dg)=P_{11}(dh)=P_{11}(gh) = 1/8 
\label{eq:outcomes_source_11}
\end{eqnarray}

\noindent
As might be expected the phase differences $\delta_1$ and $\delta_2$ play no role when the two photons are from the same source, the beam splitters just distribute the outcomes evenly to all detectors.  

\smallskip
The case of one photon from each source is more interesting.  Expanding the last line of Equation~\eqref{eq:initial_two_sky_photons} and propagating through the beam splitters, the results for the probabilities are:
\begin{eqnarray}
P_{12}(cc) & =& P_{12}(dd)=P_{12}(gg)=P_{12}(hh)  
                                 =  1/8 \nonumber \\
P_{12}(cg) & = & P_{12}(dh)  =   
       (1/8) (1+\cos(\delta_{1} - \delta_{2})) \nonumber \\
P_{12}(ch) & = & P_{12}(dg)  =   
       (1/8) (1-\cos(\delta_{1} - \delta_{2}))
\label{eq:outcomes_source_12}
\end{eqnarray}

\noindent
with the $cd$ and $gh$ outcomes having zero probability due to Hong--Ou--Mandel cancellations \cite{HOM,Nomerotski2020}, a well-known
quantum effect.  

Equation~\eqref{eq:outcomes_source_12} is the analogue of Equation~\eqref{eq:prob_one_photon}, and we can see that the relative populations of two-photon outcomes $cg$ and $dh$ versus $ch$ and $dg$ will be sensitive to the difference in the phase {\em differences} experienced by the photons on their way into the two stations.  This, in turn, is directly related to the opening angle on the sky between the two sources and so will give us access to relative astrometry information.

From this point forward we will assume that we are post-selecting on events with one photon in each station and thus consider only the $cg$, $dh$, $ch$ and $dg$ outcomes.  To count the total number of each outcome we can simply sum over the three types of pairs in Equation~\eqref{eq:initial_two_sky_photons}, since they are all mutually incoherent, e.g.
\begin{equation}
N(xy) = P_{11}(xy) N_{11} + P_{22}(xy) N_{22} + P_{12}(xy) N_{12}
\label{eq:incoherent_pair_combining}
\end{equation}

\noindent
where $x \in \{c,d\}$, $y \in \{g,h\}$ and $N_{11}$, $N_{22}$ and $N_{12}$ are the total numbers of incident pairs for the three different source combinations.  

Equation~\eqref{eq:incoherent_pair_combining} is essentially an informal version of a density matrix over the three states shown in Equation~\eqref{eq:initial_two_sky_photons}.  A more complete and rigorous derivation is carried out in Section~\ref{sec:field_theory_calc} where the radiation field is treated with a full density matrix in the basis of coherent states; and Equation~\eqref{eq:incoherent_pair_combining} can be seen as describing two-photon observable states in the limit of weak thermal radiation.

We take all the telescopes and detectors to be identical, and then make the quick semi-classical approximation that the pair counts should simply be proportional to the products of the two sources' intensities:
\begin{eqnarray}
    N_{11} & = & k \, {S_{1}}^2  \quad
    N_{22} = k \, {S_{2}}^2  \quad
    N_{12} = 2 k \, S_{1} S_{2}  \nonumber \\
    k & \equiv & \tau \; \Delta t \; (A \; \Delta \nu/h \nu)^2
    \label{eq:flux_counting}
\end{eqnarray}
where $S_{1,2}$ are the power spectral flux densities of the two sources at the wavelength of interest, $A$ is the effective collecting area of each telescope, $\Delta \nu$ is the detector optical bandwidth; and $\tau$ is the width of the time bin for correlation and $\Delta t$ is the length of integration.  With Equation~\eqref{eq:flux_counting} we can write the expectation value for the total number of each type of coincidence:
\begin{eqnarray}
\langle N(xy) \rangle & = &  
\frac{k (S_{1}+S_{2})^2}{8} 
\left[ 1 \pm  V_\mathrm{2PS} \cos (\delta_{1} - \delta_{2}) \right] 
  \nonumber \\
V_\mathrm{2PS} & \equiv & \frac{2 S_{1} S_{2}}{(S_{1}+S_{2})^2}
\label{eq:two_source_rates_pointsources}
\end{eqnarray}
\noindent
where the $+$ obtains for the $cg$ and $dh$ combinations, and the $-$ for $ch$ and $dg$; and $V_\mathrm{2PS}$ now indicates the two-point-source fringe visibility in the semi-classical approximation.

This simple formulation neglects effect of thermal photon statistics, which will boost the rate of pairs from the same source relative to Equation~\eqref{eq:flux_counting}; this behavior is the essence of Hanbury Brown \& Twiss intensity interferometry. This enhancement is, however, irrelevant to the present technique; and in fact its main consequence will be to lower the two-point-source visibility in Equation~\eqref{eq:two_source_rates_pointsources} from 1/2 to 1/3 in the symmetric $S_{1}=S_{2}$ case.  This is accounted for properly in quantum optics derivations in Section~\ref{sec:field_theory_calc}; see also Section~\ref{subsec:relation_to_HBT} for more on the relation of the present technique to HBT intensity interferometry.

An immediate feature of Equation~\eqref{eq:two_source_rates_pointsources} is that the visibility is maximized when the sources have the same brightness, e.g. $S_{1}=S_{2}$; and in the limit of extremely asymmetric sources with $S_{1} \gg S_{2}$ will fall off as the ratio $V  \propto \, S_{2}/S_{1}$.  This makes sense intuitively, that the set of all photon pairs from highly asymmetric sources will be dominated by pairs with both from the brighter source, which will then wash out the visibility by boosting the outcomes described in Equation~\eqref{eq:outcomes_source_11} over those in Equation~\eqref{eq:outcomes_source_12}. The full two-source fringe visibility as derived in Section~\ref{sec:field_theory_calc} has this same behavior, as can be seen in Equation~\eqref{Vis}.

Extending the treatment for the BSI in Section~\ref{subsec:single_source_interf} we now identify the phase differences $\delta_{1}, \delta_{2}$ each as combinations of the difference in path length down from the sky to the receivers and some path length differences within the apparatus along the legs and from the receiver to the beam splitter:
\begin{eqnarray}
\delta_{1} & = & 2 \pi (B_{ae} \sin \theta_{1} + \Delta L_{ae})/\lambda 
\nonumber \\
\delta_{2} & = & 2 \pi (B_{bf} \sin \theta_{2} + \Delta L_{bf})/\lambda 
\label{eq:delta_12_def}
\end{eqnarray}

\noindent
The angles $\theta_1$ and $\theta_2$ are the directions to the two sources, naturally generalizing the angle direction $\theta$ for the one-source case as shown in Figure~\ref{fig:single_photon_beamsplitter}.  For simplicity and compactness we will assume that the $a-e$ and $b-f$ baselines are the same $B_{ae} = B_{bf} \equiv B$; and we will combine the difference in the instrumental path length differences into a single quantity $\Delta L \equiv \Delta L_{ae} - \Delta L_{bf}$.

The pair rates from Equation~\eqref{eq:two_source_rates_pointsources} then become
\begin{widetext}
\begin{equation}
    \langle N(xy) \rangle  =   
\frac{k (S_{1}+S_{2})^2}{8} 
\left[ 1 \pm  V_\mathrm{2PS} \cos \left[ 
        \frac{2 \pi B}{\lambda} (\sin \theta_{1} - \sin \theta_{2}) +
        \frac{2 \pi \Delta L}{\lambda}
      \right] \right] 
\label{eq:pair_rates_full_pointsources}      
\end{equation}
\end{widetext}

Analogously to  Equation~\eqref{eq:Observable_BSI} for the single-photon BSI we can now define the corresponding observable for the double-source interferometer (DSI)
and write its expectation value in the case of two point sources:

\begin{eqnarray}
     O_\mathrm{DSI} & \equiv & \frac{[N(cg)+N(dh)]-[N(ch)+N(dg)]}
     {N(cg)+N(dh)+N(ch)+N(dg)}  \label{eq:Observable_DSI} \\
     \langle O_\mathrm{DSI} \rangle & = & 
     V_\mathrm{2PS} \cos \left[ 
        \frac{2 \pi B}{\lambda} (\sin \theta_{1} - \sin \theta_{2}) +
        \frac{2 \pi \Delta L}{\lambda}
      \right] \nonumber 
\end{eqnarray}

\noindent
As expected, we can see that the double-source interferometry observables $\langle N(xy) \rangle$ and $O_\mathrm{DSI}$ are directly sensitive to the {\em difference} in the sky positions between the two sources, and thus to their relative astrometry.

%
\subsection{Extended sources}
\label{subsec:extended_sources}

As mentioned above the derivation thus far assumes that the two sources are effectively point sources, i.e. that the phase difference of each arriving wavefront is perfectly the same for all photons from a given source.  We can see intuitively that this assumption will have to break down for sufficiently extended sources.  Specifically, if the angular extent of a source $\Delta\psi$ is much greater than the natural interferometric resolution $\lambda/B$ then the phase differences for a photon's arrival at the two stations will span across more than one full $2\pi$ cycle.  Thus the $\delta_{1}-\delta_{2}$ difference appearing in Equations ~\eqref{eq:outcomes_source_12} and~\eqref{eq:two_source_rates_pointsources} will be sufficiently different for different photon pairs that the observables in Equations~\eqref{eq:Observable_DSI} and~\eqref{eq:pair_rates_full_pointsources} will be significantly washed out.

The net effect of large, extended sources will be to reduce the visibility; while sources with angular sizes on the order of $\lambda/B$ will show effects on the visibility as a function of $\lambda$ and $B$ that are specific to the source's intensity profile.  This is addressed quantitatively in the full derivation in Section~\ref{sec:field_theory_calc} below.  

The essential result is that the visibility will be modified by a factor which is the product of the normalized Fourier moments of the sources' intensity distributions, projected along the $\theta$ direction, evaluated at a sky angle wave number of $2 \pi B \cos \theta / \lambda$; these appear as $\xi_{1}$ and $\xi_{2}$ in Equations~\eqref{Vis} and~\eqref{fourier-trans}.  Note that a very similar dependence appears in HBT intensity interferometry for imaging a single source, as shown in Equation~\eqref{eq:HBT_visibility} below. The magnitude of these normalized Fourier moments will go to unity for small sources, recovering the point source result, and will decrease for large sources, with some source-specific behavior in between.  We can make the general observation, as described again in Section~\ref{subsec:bright_star_example} below, that the precision of the astrometric measurement will fall off, possibly sharply, in the long baseline limit where $\Delta\psi \gg \lambda/B$.

%
\section{Quantum optics derivations}
\label{sec:field_theory_calc}

We can describe the technique presented in the previous Section from 
a more general and rigorous prospective. Let us employ a theoretical description of the fourth-order interference (or two-photon interference) phenomena, which is quite standard in quantum optics  \cite{Ou1988,Shih2003,Mandel1983}.
One may interpret that this interference occurs as a result of superposition of two-photon amplitudes \cite{shih2018,ou2007}, which represent different but yet indistinguishable alternatives corresponding to different paths of photons.  Namely, the first alternative is when one photon arrived from source 1 and was detected by the station $L$ and one photon arrived from source 2 and was detected by the station $R$; and the second is when one photon arrived from source 1 and was detected by the station $R$ and one photon arrived from source 2 and was detected by the station $L$. These two alternatives are completely indistinguishable and can be described by their two-photon amplitudes. Superposition of these amplitudes leads to the interference phenomena and allows one to reconstruct the value of $\Delta \theta$.

Generally, in quantum optics the thermal sources can be described using the Glauber-Sudarshan representation \cite{Glauber1963,Sudarshan1965} of the density operator. In our case of two stars we can write the
{\it{joint}} state of light using the following factorized form of the density operator:
\begin{eqnarray}\label{gen_state}
\hat{\varrho} 
 = \hat{\varrho}_{1}\otimes\hat{\varrho}_{2} = 
\left(\int P_{1}(\alpha_{1})\ket{\{\alpha_{1}\}}\bra{\{\alpha_{1}\}}\prod_{\omega} d^2\alpha_{1}\right)\otimes 
\left(\int P_{2}(\alpha_{2})\ket{\{\alpha_{2}\}}\bra{\{\alpha_{2}\}}\prod_{\omega}d^2\alpha_{2}\right),
\end{eqnarray}
where  $\ket{\{\alpha\}} = \prod_{\vec{k}}\ket{\alpha_{\vec{k}}}$ represents a multi-mode coherent state and the functions $P_{j}$ follow:
\begin{eqnarray}\label{Glauber_Sudarshan}
P_{j}({\alpha_{\vec{k}_{j}}}) = \prod_{\vec{k}_{j}}\frac{1}{\pi\overline{n}_{j}(\omega_{\vec{k}})}e^{-\left(\frac{|\alpha_{\vec{k}_{j}}|^2}{\overline{n}_{j}(\omega_{\vec{k}_{j}})}\right)}, \ j=1,2
\end{eqnarray}
We assumed above that
the initial state produced by each source, before transformation on the beam splitter, has well-defined spatial modes specified by wave-vectors $\vec{k}_{1}$, $\vec{k}_{2}$ and, therefore, by angles $\theta_{1}$, $\theta_{2}$.  And, the
 $\overline{n}_{1,2}(\omega_{\vec{k}_{1,2}})$ are the average numbers of photons, i.e. mean occupation number, over each frequency mode and follow a Planck distribution.  We note that in all realistic cases of observing stars the occupation numbers for any given mode will be very small, $\overline{n} \ll 1$; see \cite{Tsang2016} and also references therein.

%
\subsection{Two-photon interference}
\label{subsec:field_theory_two-photon}

To describe the observable effects in terms of joint probability for the coincident photon detection (so one photon in station L and one in R), one usually considers the propagation of electrical field operators (modes), which are carrying all information about optical paths \cite{Glauber1963}.

\begin{figure}
\begin{center}
\includegraphics[width=0.50\linewidth]{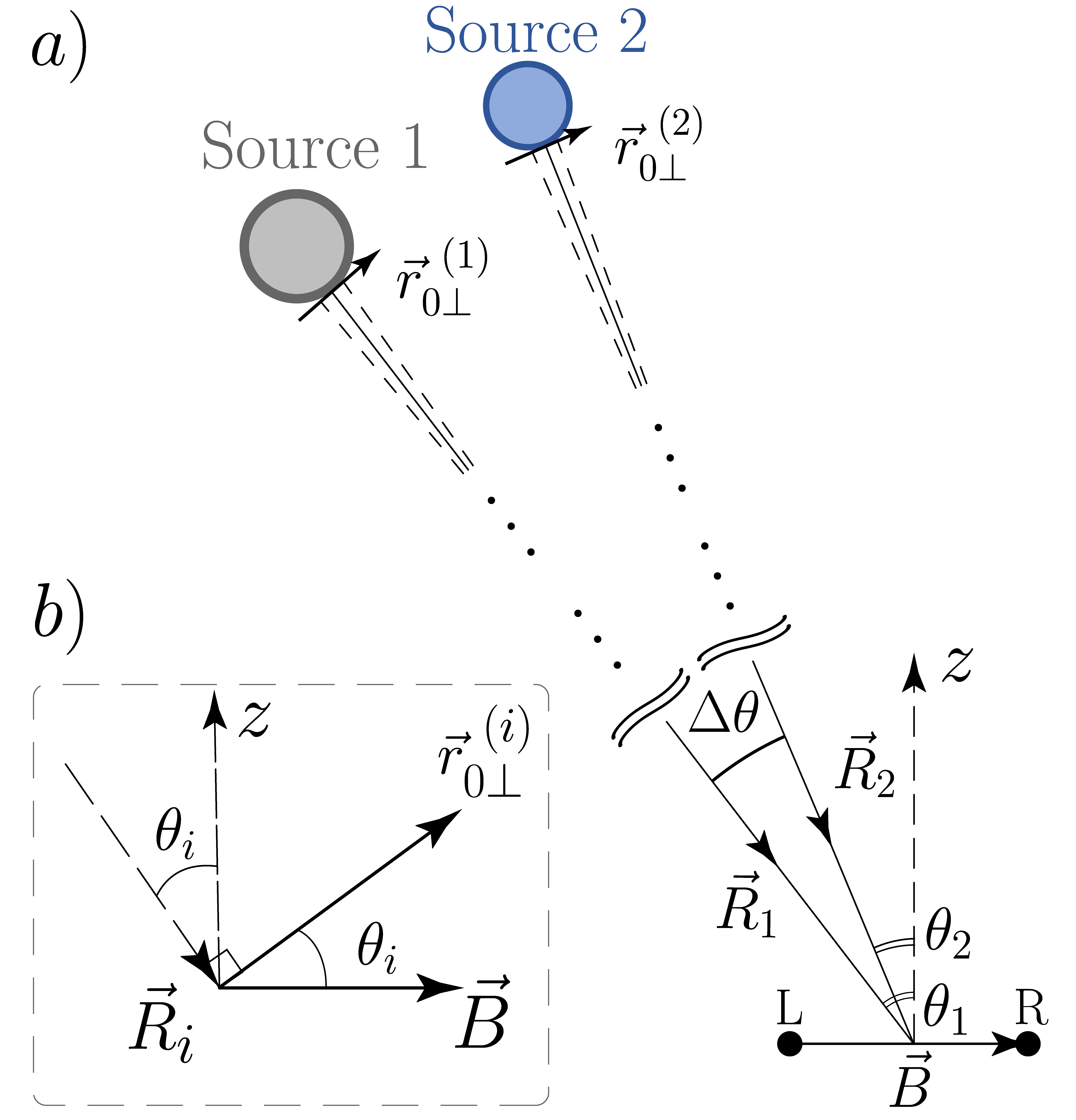}
\caption{The simplified geometry which is used to calculate the field operators and fourth-order coherence function. In the panel a) vectors $R_{1}$ and $R_{2}$ indicate the direction of incident wave vectors from both sources within far-field approximation, vector $z$ is a direction to the zenith which is orthogonal to the baseline vector $\vec{B}$, angles $\theta_{1}$ and $\theta_{2}$ are equatorial polar angles of sources relative to the axis defined by the vector $z$.
The sub panel b) illustrates the mutual arrangement of vectors presented in derivations. For given direction of $\vec{R}_{i}, i=1,2$ one can find the orthogonal projection of $\vec{B}$ to the plane of corresponding source, where each point of this plane is specified by vector $\vec{r}_{0\perp}^{i}$, we also use $D_{j}$ as effective diameter of a source. The more detailed derivation is presented in the Appendix \ref{app:field}.}
\label{fig:geometry}
\end{center}
\end{figure}

We can represent the positive frequency part of the electric field operator in the following form:
\begin{eqnarray}\label{propagation}
\hat{E}^{(+)}(\vec{r},t) = \int G_{\omega}(\vec{r}|\vec{r}_{0,\perp})c_{\omega_{\vec{k}}}\hat{a}_{\vec{k}_\perp}(\omega)e^{i\vec{k}_{\perp}\vec{r}_{0,\perp}}e^{-i\omega t}d\vec{r}_{0,\perp}
d\vec{k}_{\perp}d\omega,
\end{eqnarray}
where we used the plane-wave expansion of field operator in the source plane, and we denote $\hat{a}_{\vec{k}_\perp}(\omega)$ as the annihilation operator for the plane-wave mode with
wavevector $\vec{k}_{\perp}$ and frequency $\omega$.
The factor $c_{\omega_{k}} =  const *i\sqrt{\hbar \omega_{\vec{k}}/2 \pi}$, where the constant depends on the final choice of units and system of measurements.  We are assuming that  $\braket{\hat{E}(\vec{r},t)^{(-)}\hat{E}^{(+)}(\vec{r},t)}$ has the dimensions of intensity, and where, of course,  $\hat{E}^{(-)}(\vec{r},t)$ is the Hermitian conjugate to $\hat{E}^{(+)}$ i.e. $\hat{E}^{(-) \ \dagger} = \hat{E}^{(+)}$.

In our analysis we do not take into account the vector structure of the field by fixing the polarization. The function $G_{\omega}(\vec{r}|\vec{r}_{0,\perp})$ describes the field's free propagation, mode by mode, formed by superimposed fields from each independent, point-like sub-source.  We present the detailed derivation for the propagation of field operators in Appendix \ref{app:field}. At each point of the observation plane one can describe the resulting field operator as a superposition of two operators corresponding  to each source, see Figure \ref{fig:geometry}.  We note that the formalism employed below is directly related to the van Cittert-Zernike theorem~\cite{mandel_wolf1995}. However, in our approach we are able to distinguish two different spatial modes $\vec{k}_{1}$ and $\vec{k}_2$ corresponding to sources $1$ and $2$, and, respectively, their electric field operators $\hat{E}^{{[1]}{(+)}}$ and $\hat{E}^{{[2]}{(+)}}$. Each mode in collected in each station and fed to the beam-splitter~(BS) input port, as described in Section~\ref{sec:basics_two_photon_amplitude} above. Thus, one can describe the output field operators after the BS transformation through the input operators, see Appendix \ref{app:field} for details:
\begin{eqnarray}\label{field_operator_init}
\hat{E}^{(+)}_{a_{s}} = \frac{1}{\sqrt{2}}\left(\hat{E}^{{[1]}{(+)}}_{s} + (-1)^{a_{s}}\hat{E}^{{[2]}{(+)}}_{s}\right) 
a_{s}\in \{0,1\}, \ \forall s = {\rm{L,R}} 
\end{eqnarray}
where index $ s =\rm{L,R}$ is referred to the observing stations and index ${a}_{s} = 0,1$ parameterizes output ports of both BS's: 
$ \{ {a = 1}_{L},{a =2}_{L} \} \equiv  \{ c,d \}$ 
and 
$ \{ {a = 1}_{R},{a =2}_{R} \} \equiv \{ g,h \}$. 

To observe the two-photon quantum interference effects we calculate the fourth-order coherence function $\Gamma_{1,2}^{a_{L},a_{R}}(\tau)$, which determines the rate of coincidences between pairs of detectors placed at the BS output ports, and we denote the time difference between detections here as $\tau$.
In our observation scheme, according to \eqref{field_operator_init}, $\Gamma_{1,2}^{a_{L},a_{R}}(\tau)$ will equal:
\begin{eqnarray}\label{corr.avarage}
\Gamma_{1,2}^{a_{L},a_{R}}(\tau) &=& \braket{\hat{E}^{{(-)}}_{a_{L}}(t)\hat{E}^{{(-)}}_{a_{R}}(t + \tau)\hat{E}^{{(+)}}_{a_{R}}(t+\tau)\hat{E}^{{(+)}}_{a_{L}}(t)} = \nonumber \\
 &=& \braket{\mathcal{T}:\hat{I}^{[1]}_{L}\hat{I}^{[1]}_{R}:}+\braket{\mathcal{T}:\hat{I}^{[2]}_{L}\hat{I}^{[2]}_{R}:}
+ \braket{\mathcal{T}:\hat{I}^{[1]}_{L}\hat{I}^{[2]}_{R}:}+\braket{\mathcal{T}:\hat{I}^{[2]}_{L}\hat{I}^{[1]}_{R}:} 
+\nonumber \\
&+& \left(-1\right)^{a_{L}+a_{R}}  \bigg[\braket{\mathcal{T}\hat{E}^{{[1]}{(-)}}_{R}\hat{E}^{{[2]}{(-)}}_{L}\hat{E}^{{[2]}{(+)}}_{R}\hat{E}^{{[1]}{(+)}}_{L}} + {\rm{c.c.}}\bigg],
\end{eqnarray}
where $\braket{\dots} = {\rm{tr}}\left(\hat{\varrho} \dots\right)$ is averaged over the ensemble of quantum states defined in \eqref{gen_state} and \eqref{Glauber_Sudarshan}. (Note that we sometimes suppress writing the dependence on time explicitly, to keep equations uncluttered.)  Here we employ the intensity operator $\hat{I}^{j}_{s}= \hat{E}^{[j](-)}_{s}\hat{E}^{[j](+)}_{s}$ with $s = \{L,R\}$ and $j=\{1,2\}$, while the symbols $\mathcal{T}$ and $::$  indicate that all operators inside expressions of the form  $\braket{\mathcal{T}:_{\dots}:}$  must be time and normal ordered~\cite{mandel_wolf1995}.  This is the quantum analog of the four-point correlator of classical fields as seen in Equation~\eqref{eq:four_point_classical} below.

Let us assume a quasi-monochromatic approximation ($\Delta\omega \ll \omega_{0}$), appropriate for a very narrow filter, say a bandwidth on the order of $\sim 1 {\rm{GHz}}$ as imagined in Section~\ref{subsec:bright_star_example} below.  One then needs to substitute the explicit form of expressions~\eqref{field_operator_init}  and~\eqref{propagation} (see Equation~\eqref{field_operator_init-a} in Appendix \ref{app:field}) to calculate~\eqref{corr.avarage}.
Under these assumptions a somewhat long but straightforward calculation yields: 
\begin{eqnarray}\label{corr.rewritten}
\Gamma_{1,2}^{a_{l},a_{r}}(\tau) &\approx& I_{1}^2(1+|\gamma_{1}(\omega_{0},\tau)|^2) + I_{2}^2(1+|\gamma_{2}(\omega_{0},\tau)|^2) + \nonumber \\ 
&+&2I_{1}I_{2}\bigg[1 + (-1)^{(a_{L}+a_{r})}|\gamma_{1}(\omega_{0},\tau)||\gamma_{2}(\omega_{0},\tau)|
\cos\left(\frac{\omega_{0}B}{c}(\sin\theta_{1} - \sin\theta_{2}) + \frac{\omega_{0}\Delta L}{c}\right)\bigg]
\end{eqnarray}
where $I_{1}$ and $I_{2}$ are the average intensities of the sources, defined by: 
\begin{eqnarray}\label{eq:intensities_def}
I_{j} = \frac{\hbar\omega_{0}^3}{8\pi c^2 R_{j}^2}{\rm{FT}}_{\Sigma_{j}}\left(0\right)\overline{N}_{j}(\omega_{0}) 
=\frac{1}{16}\hbar \omega_{0}\left(\frac{\omega_{0}D_{j}}{cR_{j}}\right)^2\overline{N}_{j}(\omega_{0}), \ j=1,2
\end{eqnarray}
with the coordinates $R_{j}$ and $D_{j}$ following the geometry laid out in Figure~\ref{fig:geometry}.

In~\eqref{eq:intensities_def} we denoted $\overline{N}_{j}(\omega_{0})$ as an average photon flux after Gaussian filtration in the frequency domain by a filter with a narrow bandwidth $\Delta \omega$ and central frequency $\omega_{0}$:
\begin{eqnarray}\label{intensity}
&&\overline{N}_{j}(\omega_{0}) = \int  \overline{n}_{j}(\omega)\exp{\left(-\frac{(\omega - \omega_{0})^2}{2\Delta \omega^2}\right)}d\omega
\approx 
\sqrt{2\pi}\overline{n}_{j}(\omega_{0})\Delta\omega. 
\end{eqnarray}
The symbol $\rm{FT}_{\Sigma_{j}}\left(\cdot\right)$ above stands for Fourier image of the intensity distribution of source $j$.  In the general case we can write this as follows:
\begin{eqnarray}\label{FT-spatial}
{\rm{FT}}_{\Sigma_{j}}\left(\frac{\omega\vec{B}\cdot\vec{r}_{0\perp}^{j}}{c}\right) = \int_{\Sigma_{j}}{\rm{exp}}\left({i\frac{\omega}{c}\vec{B}\cdot\vec{r}_{0\perp}^{j}}\right)d\vec{r}_{0\perp}^{(j)};\ j=1,2;
\end{eqnarray}
The Fourier transform is taken over the area of one source and we introduce $\Sigma_{j}$, as a characteristic area of the source projection on the object plane; in the case of sharp-edge disk model for example $\Sigma_{j} = \pi D_{j}^2/4$.

Within the quasi monochromatic approximation the auto-correlation functions $\gamma_{1,2}(\omega_{0},\tau)$ are given by: 
\begin{eqnarray}
\gamma_{j}(\omega_{0}, \tau) = \frac{\braket{\hat{E}^{{[j]}{(-)}}_{L}\hat{E}^{{[j]}{(+)}}_{R}}}{I_{j}} \approx 
\frac{1}{\Sigma_{j}}{\rm{FT}}_{\Sigma_{j}}\left(\frac{\omega\vec{B}\cdot\vec{r}_{0\perp}^{j}}{c}\right)
{\rm{FT}}_{{\rm{filter}}(\omega_{0})}\left(\frac{B\sin(\theta_{j})}{c} - \tau+\delta_{jR} -\delta_{jL}\right).
\end{eqnarray}
Here, we introduced ${\rm{FT}}_{{\rm{filter}}}$ similarly to \eqref{FT-spatial} - the Fourier transform of Gaussian filter function in the frequency domain:
\begin{eqnarray}\label{FT - freq.}
&& {\rm{FT}}_{{\rm{filter}}(\omega_{0})}\left(\frac{B\sin(\theta_{j})}{c} - \tau+\delta_{jR} -\delta_{jL}\right) = 
\int\frac{\exp{\left(i\omega\left(\frac{B\sin{\theta_{j}}}{c} - \tau +  \delta_{jR} - \delta_{j L}\right) -\frac{(\omega - \omega_{0})^2}{2\Delta \omega^2}\right)}}{\sqrt{2\pi\Delta\omega^2}}d\omega = \nonumber \\
&&\exp{\bigg[-\left(\frac{B\sin(\theta_{j})}{c} - \tau+\delta_{jR} -\delta_{jL}\right)^2\frac{\Delta\omega^2}{2}\bigg]} 
\exp{\bigg[i\omega_0\left(\frac{B\sin(\theta_{j})}{c} - \tau+ \delta_{jR} -\delta_{jL}\right)\bigg]}
\end{eqnarray}
To simplify further the above expressions let us
denote $\xi_{j}\left(\frac{\omega_{0} B}{c} \cos \theta_{j}\right)$ 
as the normalized Fourier coefficient of the source distribution:
\begin{eqnarray}\label{fourier-trans}
\xi_{j}\left(\frac{\omega_{0} B}{c} \cos \theta_{j}\right) = \frac{1}{\Sigma_{j}}{\rm{FT}}_{\Sigma_{j}}\left(\frac{\omega\vec{B}\cdot\vec{r}_{0\perp}^{j}}{c}\right) =
\frac{ 2J_{1}\left(\frac{\omega_{0}}{2c R_{j}}BD_{j}\cos{\theta_{j}}\right)}{\frac{\omega_{0}}{2c R_{j}}BD_{j}\cos{\theta_{j}}}, j=1,2
\end{eqnarray}
The second line in~\eqref{fourier-trans} is the result of explicitly integrating over $\vec{r}_{0\perp}$ assuming a sharp-edge disc model for each source; $J_{1}(x)$ is a Bessel function of the first kind.  Similar to Equation \eqref{intensity} integration over $d\omega$ in the case of a Gaussian frequency filter gives:
\begin{eqnarray}\label{corr_xi}
\gamma_{j}(\omega_{0}, \tau) \approx \xi_{j}\left(\frac{\omega_{0} B}{c} \cos \theta_{j}\right) 
{\rm{FT}}_{{\rm{filter}}(\omega_{0})}\left(\frac{B\sin(\theta_{j})}{c} - \tau+\delta_{jR} -\delta_{jL}\right)
\label{eq:vis_factors}
\end{eqnarray}
where we have pulled out of the integral all functions which are  changing slowly with time, compared to the quickly oscillating complex exponential function in the integral.

Note that one can rewrite the phase delays as $\delta_{1R} - \delta_{1L}=\Delta L_{ae}/c$, $\delta_{2R} - \delta_{2L}  = \Delta L_{bf}/c$, and $(\Delta L_{ae} - \Delta L_{bf})/c =  \Delta L/c$ in accordance with notations introduced earlier in Equations~\eqref{eq:delta_12_def} and~\eqref{eq:pair_rates_full_pointsources}.
Also note that though we have treated our source as a sharp disc, it is possible to use an arbitrary source model composed of sub-sources, which all have Gaussian spatial distributions. In this case one obtains the functions $\gamma^{G}_{1,2}(\omega_{0},\tau)$ by a simple replacement of  $\frac{J_{1}(x_{1,2})}{x_{1,2}}$  in Equation \eqref{fourier-trans} with ${\rm{exp}}\left(-x_{1,2}^2
\right)$, where $x_{1,2} = \frac{\omega_{0}}{c R_{1,2}}BD_{1,2}\cos{\theta_{1,2}}$.

%
In accordance with Glauber's photodetection theory \cite{Glauber1963} the differential joint probability  $P^{\rm{two \ photons}}_{L,R, \tau}$ to observe the coincident events is proportional to the fourth-order coherence function $P^{\rm{two \ photons}}_{L,R, \tau} \sim \Gamma_{1,2}^{a_{l},a_{r}}(\tau) $ laid out in Equation~\eqref{corr.avarage}
and expressed explicitly in 
Equation~\eqref{corr.rewritten}. Thus, one can estimate counting rate of the coincidence, e.g. expected pair counts, within a short, single-bin, time interval $T_{r}$ corresponding to the resolution time of detectors:
\begin{eqnarray}\label{coincedence}
N_{c}(xy) &=&\int_{-T_{r}/2}^{T_{r}/2}P^{\rm{two \ photons}}_{L,R, \tau}d\tau = \nonumber \\
& = & A^2\eta_{1}\eta_{2}T_{r} \bigg[(I_{1}+I_{2})^2 + I_{1}^2\frac{T_{c}g_{11}}{T_{r}} + I_{2}^2\frac{T_{c}g_{22}}{T_{r}} \pm
2I_{1}I_{2}\frac{T_{c}g_{12}}{T_{r}}\cos\left(\frac{\omega_{0}B(\sin\theta_{1} - \sin\theta_{2})}{c} + \frac{\omega_{0}\Delta L}{c}\right)\bigg]
\end{eqnarray}
where the $+$ obtains for the $xy=cg,dh$ combinations, and the $-$ for $xy=ch,dg$. 
We denote as $T_{c} \sim 1/\Delta \omega$ a {\it{characteristic coherence time}} after filtering, $\eta_{1,2}$ are the detector quantum efficiencies and $A$ is the effective collection area. Note, for clarity we replaced $a_{L}\rightarrow x, a_{R}\rightarrow  y$  in Equation~\eqref{coincedence} using the labelling notation in Section~\ref{sec:basics_two_photon_amplitude}. This is now the fully rigorous generalization of Equation~\eqref{eq:pair_rates_full_pointsources} for extended, thermal sources.

The functions $g_{ij},i,j = 1,2$ in Equation~\eqref{coincedence} are defined as follows:
\begin{eqnarray}\label{f-rates}
&&g_{ij} = \frac{1}{T_{c}}\int_{-T_{r}/2}^{T_{r}/2}|\gamma_{i}(\omega_{0},\tau)||\gamma_{j}(\omega_{0},\tau)|d\tau \nonumber \\
&& \ \frac{T_{c}}{T_{r}} g_{ij}\propto \xi_{i}\left(\frac{\omega_{0} B}{c} \cos \theta_{i}\right)\xi_{j}\left(\frac{\omega_{0} B}{c} \cos \theta_{j}\right), \Delta\omega T_{r} \sim T_{r}/T_{c} \ll 1, \,\,\, i,j = 1,2.
\end{eqnarray}

Equation~\eqref{coincedence} connects the two-photon count rates after interference to information on the two sources' relative positions; we can re-cast
the observed pair coincidences in the following form:
\begin{eqnarray}\label{coince_2}
&&N_{c}(xy)\propto \big[1 \pm V \cos\left(\frac{\omega_{0}B(\sin\theta_{1} - \sin\theta_{2})}{c} + \frac{\omega_{0}\Delta L}{c}\right)\big], \nonumber \\
&& \ 
\end{eqnarray}
where $V$ is the fringe visibility, also discussed in detail later in Section \ref{sec:sky_observables}. In the case of very narrow frequency filter (as we assumed $\Delta \omega T_{r} \ll 1$), we can write the fringe visibility $V$ in accordance with \eqref{coincedence}, \eqref{f-rates} and \eqref{fourier-trans} as follows:
\begin{eqnarray}\label{Vis}
V =
\frac{2I_{1}I_{2}\xi_{1}\xi_{2}}{(I_{1}+I_{2})^2 + (I_{1}\xi_{1})^2 + (I_{2}\xi_{2})^2},
\end{eqnarray}
where we put $\xi_{j} \equiv \xi_{j}\left(\frac{\omega_{0} B}{c} \cos \theta_{j}\right)$ for simplicity.  Equation~\eqref{Vis} is now the full visibility including extended sources and all quantum effects, generalizing the semi-classical, two-point-source visibility described earlier in Section~\ref{sec:basics_two_photon_amplitude}.

\medskip

We can gain some insight into the essential features of the two photon interference at work by considering a "toy" model where hypothetical sources produce a state over just two modes, which can be modelled by the very simple density operator: 
$$\hat{\varrho}_{11} \propto (1-\overline{n}_{1}\overline{n}_{2})\ket{0_{1}0_{2}}\bra{0_{1}0_{2}}+\overline{n}_{1}\overline{n}_{2}\ket{1_{1}1_{2}}\bra{1_{1}1_{2}}$$
We can then rewrite the differential joint probability corresponding to the detection of one photon at the station $L$ and one photon at the station $R$: 
\begin{eqnarray}\label{reduced prob.}
P^{\rm{two \ photons}}_{L,R, \tau} \, \propto \, \overline{n}_{1}\overline{n}_{2}|\Psi_{1,2}(L,R) + (-1)^{a_{L}+a_{R}}\Psi_{1,2}(R,L)|^2,
\end{eqnarray}
where we used $\Psi_{1,2}(L,R)$ and  $\Psi_{1,2}(R,L)$ to denote the effective two-photon amplitudes. Following \cite{Shih2003,shih2018} they can also be termed two-photon effective wavefunctions: 
\begin{eqnarray}\label{wavefuntion}
\Psi_{1,2}(s1,s2) = \bra{0_{1}0_{2}}\hat{E}^{{[1]}{(+)}}_{s1}\hat{E}^{{[2]}{(+)}}_{s2}\ket{1_{1}1_{2}}, s_{i} = L,R \  \forall i = 1,2
\end{eqnarray}
Equations~\eqref{reduced prob.} and~\eqref{wavefuntion} express the nature of two-photon interference, as described at the beginning of this section.

Moreover, one can establish the direct analogy of basic idea presented in Section~\ref{sec:basics_two_photon_amplitude} based on the entanglement (see caption of 
Figure~\ref{fig:interferometer_spread_detectors_wavefront}) and interference between the two-photon amplitudes. The analogy stems from the non-factorizable and non-local behavior of the two-photon interference, which are also characteristic properties of the entanglement.

%
\subsection{Relation to Hanbury Brown \& Twiss intensity interferometry}
\label{subsec:relation_to_HBT}

It will be recognized that the two-photon technique described here is somewhat akin to the celebrated Hanbury Brown \& Twiss~(HBT) intensity correlation technique~\cite{hbt, Foellmi2009},  particularly since (i)~both require detection of two photons from the sky, and (ii)~both accomplish optical interferometry using two independent optical systems, requiring only a classical signal link between them.  However, there are important differences in the information that can be accessed in the two techniques, which are worth highlighting here. 

\begin{figure}
\begin{center}
\includegraphics[width=0.85\linewidth]{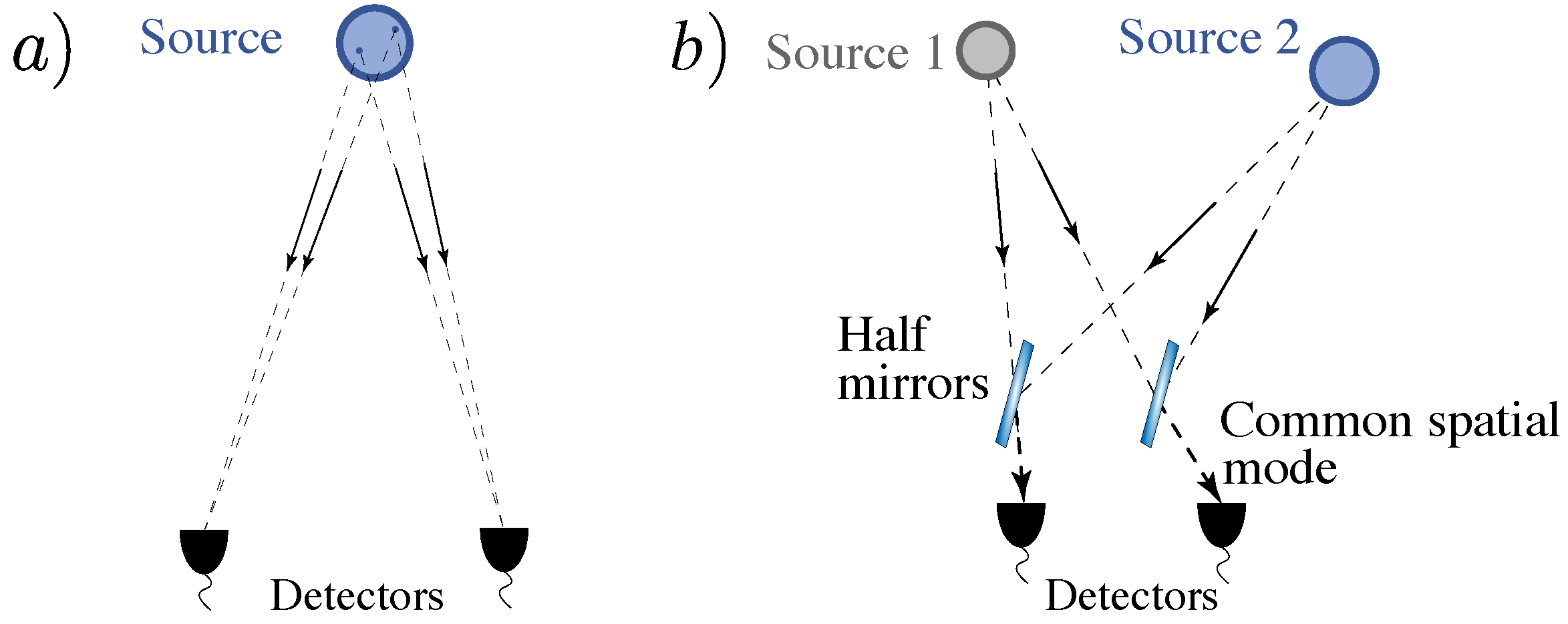}
\caption{Generalization of (a) standard HBT intensity interferometry to (b) two-photon amplitude interfereometry from two sky sources; see text for detailed discussion.}
\label{fig:HBT_generalization}
\end{center}
\end{figure}

A standard astronomical HBT measurement is illustrated in Figure~\ref{fig:HBT_generalization}a.  There are two photons produced from different points on an extended source, and each impinges on both detector apertures exciting a mode similar to that described in Equation~\eqref{eq:initial_single_sky_photon}.  If the separation between the two source points is small enough so that both photons are in essentially the same spatial mode, e.g. $\Delta \theta < \lambda/B$, and the source follows thermal photon statistics, then the rate for coincident detections at the two stations will be enhanced relative to the bare combinatoric rate.  For larger sources or longer baseline separations, e.g. $B \; \Delta \theta > \lambda$,  the enhancement will fall off, thus providing sensitivity to the source size via baseline scan.

Astronomical HBT interferometry has historically been used mainly to measure the gross properties of single objects, e.g. stellar diameters and limb darkening.  It was realized, though, that if the source is not a single object but two point sources, as in a close binary, then intensity interferometry can yield some information on the sources' separation; see~\cite{HanburyBrown1974} for discussion and results, and~\cite{Twiss1969} for useful formulae.  However, as noted by Twiss~\cite{Twiss1969} this is only possible for very small opening angles between the two sources, on the order of arc-seconds or less, where both can excite the same mode going into a single aperture.  Thus, standard HBT can never be used for astrometry other than within a single compact binary, and there the observable signal is not very distinct.

Now imagine the scheme illustrated in Figure~\ref{fig:HBT_generalization}b: here we have taken the light from two sources and combined them into the same spatial modes arriving on the detector apertures.  We can see that this is quantum optically equivalent to the proposed arrangement as shown in Figure~\ref{fig:interferometer_spread_detectors_wavefront}, if we use only two of the four final detectors, since the half mirrors are functionally beam splitters.  The two photons shown, e.g. gray and blue, will now interfere with each other in the detectors, yielding the oscillatory observable with  direct dependence on the pair opening angle as shown in Equations~\eqref{eq:Observable_DSI} and~\eqref{eq:nu_fringe_smallopen}.  Thus it is not unreasonable to describe the present technique as an extension or generalization of HBT interfereometry.  But an important difference is that now the effects of differences in relative phase difference are detectable, and even amplified, which is not a feature of intensity interferometry.  At the same time, the pair correlation enhancement at the core of the HBT technique is specific to thermal sources, while the new oscillatory observable would be seen with any pair of sources, thermal, coherent or single-photon. We can thus term the present technique as ``two-photon amplitude interferometry'' to distinguish it clearly from simply a new version of intensity interferometry.

We can see the generalization in mathematical terms, first in
the classical picture discussed in Section~\ref{sec:astrometry_interferometry} and Appendix~\ref{app:classical} where the connection can be understood quite intuitively. Both approaches involve four-point correlators (as opposed to two-point in the normal amplitude interferometry); but in the HBT scheme all four correlated quantities come from the same source, while here they come from two different sources. The crucial difference is that in the HBT effect the only quadratic quantity one can form from the single source corresponds to the intensity, in which the phase information is lost. In the two-source interferometry, the single telescope quadratic quantity is the signal cross-correlation, which is a complex quantity, thus encoding a phase difference between the two sources.

Consequently, the HBT observables measure the magnitude of Fourier transformation of the image plane, while the two-source interferometry measures the conjugate product of Fourier transforms of two image planes. In this case the phase information is preserved fully if one of the sources is a point source.

In the quantum picture, the differences can be clearly seen by examining Equation~\eqref{corr.rewritten}. The first two terms, namely $I_{1}^2(1+|\gamma_{1}(\omega_{0},\tau)|^2)$ and $I_{2}^2(1+|\gamma_{2}(\omega_{0},\tau)|^2)$, represent the correlated intensity fluctuations (as in the HBT effect) in the case of each source, independently. In the case of two point-like sources in the same field of view  with well  distinguished $\vec{k}_{1}$ and $\vec{k}_{2}$, we end up with expressions similar to \eqref{corr.rewritten}, including the oscillatory term similar to $~ \cos\left(\frac{\omega_{0}B}{c}(\sin\theta_{1} - \sin\theta_{2}) + \frac{\omega_{0}\Delta L}{c}\right)$ as was determined, for example, in \cite{Twiss1969}.

In contrast, in the presented approach, transformation of the field operators (or, equivalently, of the photon states) allows one to extract the angular information encoded in optical paths with higher accuracy. Again, this is only possible because we assume that we are able to distinguish and effectively collect the spatial modes from each source and operate with states like $\hat{\varrho}_{\vec{k}_{1}}\otimes\hat{\varrho}_{\vec{k}_{2}}$. Thus, the proposed amplitude interferometry captures more information from the incident photon field and can be seen as essentially a generalization of the original HBT technique.

%
\section{Sky observables}
\label{sec:sky_observables}

The usual goal of traditional interferometry is {\em imaging}, ie reconstructing the shape and size of a source's brightness distribution on the sky.  In image reconstruction the primary observable for an observation from a given pair of receivers is the amplitude, e.g. visibility, and phase of the interference fringe pattern.

For astrometry, however, we are interested in the relative positions of different sources, and we can access this in the present scheme by instead observing the {\em spacing} of the fringes for the observables in Equations~\eqref{eq:pair_rates_full_pointsources} and~\eqref{eq:Observable_DSI} during interferometric observations.

%
\subsection{Earth rotation fringe rate}
\label{subsec:drift_scan_fringe_rate}

We can illustrate the essential idea by imagining an idealized observation, where the baseline between the two stations is straight east-west and both sources lie on the celestial equator. 
The path differences will then be gradually modulated by Earth's rotation. We can write the source position angles $\theta_{1}$ and $\theta_2$ as
functions of time
\begin{equation}
    \theta_{1}(t) = \theta_{0} + \Omega_\Earth t
    \qquad
    \theta_{2}(t) = \theta_{1}(t) + \Delta \theta
\end{equation}
\noindent
where $\theta_{0}$ is the position of source 1 at the epoch chosen as $t=0$, $\Delta \theta$ is the opening angle between the sources, and $\Omega_\Earth$=1.16$\times$10$^{-5}$rad/sec is the angular velocity of the Earth's rotation.  Substituting into Equation~\eqref{coince_2} and then expanding to first order in $\Omega_\Earth t \ll 1$  we can now write the average number of observed pair coincidences as a function of time with four parameters:
\begin{equation}
    \langle N_{xy} \rangle (t) =
    \bar{N}_{xy} \left[ 1 \pm  V \cos \left(\omega_{f} t + \Phi \right) \right]
    \label{eq:basic_time_dependence}
\end{equation}
\noindent
Here we use $\bar{N}_{xy}$ for the average observed number of pairs of type $xy$, with the ``$+$'' and ``$-$'' corresponding to the different pair types, e.g. $cg$,$dh$ versus $ch$,$dg$;  $V$ is the fringe visibility;
and $\Phi$ is an overall phase offset reflecting the delays in the system and the value of $\theta_{0}$. The fringe angular rate $\omega_{f}$ is 
\begin{equation}
\omega_{f} = \frac{2 \pi B \Omega_\Earth}{\lambda} 
(\sin \theta_{0} \sin \Delta\theta + \cos \theta_{0}
  (1-\cos\Delta\theta))
  \label{eq:nu_fringe_full}
\end{equation}
which provides a direct measure of $\Delta\theta$ if all the other parameters are known.
In the limit of small opening angle $\Delta\theta \ll 1$ the fringe rate simplifies to
\begin{equation}
\omega_{f} = \frac{2 \pi B \Omega_\Earth \sin \theta_{0}  }{\lambda} \Delta\theta
  \label{eq:nu_fringe_smallopen}
\end{equation}
and we will use this form for simplicity.

Generally, measurements of frequency across a time domain are among the most precise; and here measurement of the fringe rate provides direct access to the opening angle. 
From this we can outline a program for dynamic astrometry.  We can make a measurement of $\omega_{f}$ every day at the same sidereal time, i.e. the same $\theta_{0}$; and then day-by-day changes in $\omega_{f}$ over a season would provide information on the evolution of $\Delta \theta$ due to parallax, orbital motions, gravitational lensing, etc., as well as relative overall proper motion.  
Quantitative estimates for precision on $\omega_{f}$ and $\Delta\theta$
follow in Section~\ref{subsec:precision} 
and we discuss a nominal example using bright stars in Section~\ref{subsec:bright_star_example}.

%
\subsection{Precision on parameters}
\label{subsec:precision}

Without describing a particular instrument we can picture the essential data stream as simply the numbers $N_{xy}$ of coincident pairs observed in the two stations $L,R$ of the four different types $\{xy\} \in \{cg,ch,dg,dh\}$ in each successive small time interval length $\Delta t$.  Assuming the binning time is small compared to the fringe period $\Delta t \ll 1/\omega_{f}$ then we look at the pair rate $n_{xy}(t)=N_{xy}/\Delta t$ as our main observables, each.  With an assumption about the statistics of the pair counts we can then fit the observable streams $n_{xy}(t)$ with functions of the type in Equation~\eqref{eq:basic_time_dependence}, each using the same four parameters.  The relevant result of each overall observing session is then an estimate of the fringe rate $\omega_{f}$, and the rest of the parameters.  We now estimate what the uncertainty on the parameters from one scan of length $T$ will be.

We use the Fisher matrix formalism, which gives the expected sensitivity for optimal estimators.  The basic quantity is the Fisher matrix, which is the expectation value of second derivative of log likelihood of a given optimal fit
\begin{equation}
    F_{ij} = \left< \frac{\partial^2 \log \mathcal{L}}{\partial \theta_i \partial \theta_j }\right>.
\end{equation}
The average is over possible realizations of the data assuming they are as given by fiducial theory. 
The \emph{marginalized} error on parameter $\theta_i$ is then given by 
\begin{equation}
    \sigma \left[\theta_i\right] = \sqrt{(F^{-1})_{ii}}
\end{equation}
\newcommand{\nbar}{\bar{n}}
\newcommand{\dtheta}{{\Delta \theta}}
\newcommand{\dL}{{\Delta L}}
During the interferometric observation scan
the pair count rates on the $cg$, $dh$, $ch$ and $dg$
detector combinations are modelled with the functions
\begin{eqnarray}
    n_{cg} + n_{dh} &=& \frac{\nbar}{4} \left[1+V \cos\left(\omega_{f} t + \Phi \right) \right], \label{eq:f1} \\
    n_{ch} + n_{dg} &=& \frac{\nbar}{4} \left[1-V \cos\left(\omega_{f} t + \Phi \right) \right], \label{eq:f2}
\end{eqnarray}
where $\nbar$ is the average rate of pairs from both sources. This is the equivalent of Equations~\eqref{eq:outcomes_source_12},  expressed as the rate of coincidences $\nbar$, which has units of inverse time.
The non-signal pairs coming from the same source as in Equation~\eqref{eq:outcomes_source_11}, or other uncorrelated sources of noise such as dark currents, etc, can be absorbed in a reduced value of visibility $V$. 

The details of the Fisher matrix calculation are given in Appendix~\ref{app:sensitivity_estimate}, where it is assumed that the sampling time interval length $\Delta t$ is short enough that the pair counts will follow a Poisson distribution.  The result in Equation~\eqref{eq:variance_omega_f} is that the standard deviation for the estimate on $\omega_{f}$ is
\begin{equation}
  \sigma \left[ \omega_f  \right] = \frac{2 \sqrt{6}}{V T \sqrt{\nbar T \kappa(V)}}
    \label{eq:stdv_omega_f}
\end{equation}
where $\kappa(V)$ is a small dimensionless auxiliary function with a value between $1/2$ and $1$, defined in Equation~\eqref{eq:aux_func_S}.

In the idealized case where $B$, $\lambda$, $\Omega_\Earth$ and $\theta_0$ are fixed we can re-write the dimensionless fractional errors in a very intuitive way
\begin{equation}
    \frac{\sigma \left[ \omega_{f} \right]}{\omega_{f}} =
    \frac{\sigma \left[ \Delta \theta \right]}{\Delta\theta} =
\sqrt{\frac{6}{\pi^{2} \kappa}} \; \frac{1}{V \, N^{Cycle} \sqrt{N^{Pair}}}
\label{eq:stdv_dimensionless}
\end{equation}
With a prefactor of close order unity, the fractional uncertainty on $\omega_{f}$ and on $\Delta\theta$ depends inversely on the three dimensionless quantities: (i)~the fringe amplitude/visibility $V$;
(ii)~the number of full fringes cycles $N^{Cycle}=T \omega_{f}/2\pi$ that pass during the observation time $T$; and (iii)~the square root of the total number of observed pairs $N^{Pair}=\nbar T$.  

Experimentally we can write the uncertainty on $\Delta\theta$ to determine our sensitivity to astrometric changes between observations on different days:
\begin{equation}
    \sigma \left[ \Delta \theta \right] =
\sqrt{\frac{6}{\pi^{2} \kappa}} \; \frac{1}{V} \, 
\frac{\lambda}{B} \, 
\frac{1}{ T \Omega_\Earth \, \sin \theta_{0} } \,
\frac{1}{\sqrt{\nbar T}}
\label{eq:stdv_deltatheta}
\end{equation}
We will note three quick observations from Equation~\eqref{eq:stdv_deltatheta}, and then move on to a quantitative evaluation: (i)~the uncertainty on $\Delta\theta$ is independent of $\Delta\theta$ itself, affording flexibility in choosing source pairs; (ii)~the uncertainty on $\Delta\theta$ goes with $\lambda/B$, allowing us to gain from longer baselines as long as the visibility is not reduced (see below); and (iii)~the overall $T^{-3/2}$ dependence on the length of the observation period is much faster than simply the $T^{-1/2}$ gain from photon pair statistics, reflecting the advantage of being able to use the measurement of a rate.

%
\subsection{Bright star example}
\label{subsec:bright_star_example}

To estimate the general magnitude of the precision that can be reached on an opening angle measurement we will model a simplified experiment using rounded but reasonable numbers; for example let us assume $\lambda=1\mu$m, and $T=10^{4}$~sec for a one-night observation; and set $\sin \theta_{0}=1/\sqrt{2}$ generically.

The choice of baseline is an optimization for a given target pair based on their angular diameters, which we will refer to as $\Delta\psi$ and assume is the same for both stars.  As long as $\lambda/B \gg \Delta\psi$ the stars can be considered point-like and the visibility $V$ will be independent of baseline, and so in this limit a longer baseline will always improve the resolution on $\Delta\theta$ as per Equation~\eqref{eq:stdv_deltatheta}.  However, in the long-baseline limit that $\lambda/B \ll \Delta\psi$ the visibility will be reduced, as mentioned in Section~\ref{subsec:extended_sources} and seen in Equation~\eqref{Vis}, eroding the precision on $\Delta\theta$ faster than the longer baseline improves it.
As such there will be an optimum baseline for the measurement of $\Delta\theta$ for any particular pair of sources.  The exact value will depend on the details of the extended source distributions, but for present purposes we will approximate the optimum as simply satisfying $\lambda/B=\Delta\psi$.

Our worked example will be for bright, high-temperature stars, which will be a reasonable starting point for a first experiment.  We will assume stars of magnitude~2, and with apparent angular diameters\footnote{An existing close pair with approximately these properties, and which lies on the celestial horizon, would be $\varepsilon$~Orionis and $\delta$~Ori~Aa1, for example.} of $\Delta\psi=$0.5~mas leading to an optimal baseline of $B=200$m.  To allow for pair brightness asymmetry, and also for the effects of finite stellar sizes, we will assume a visibility of $V=0.20$ and approximate $\kappa(V)=1/2$.

Lastly we need to estimate the rate of pairs which will be captured in the telescopes and detected in coincidence.  Independent of any exact instrument design the two basic figures of merit are the effective collecting area into each telescope and the time resolution of each single-photon detector.  
We assume the aperture, collection and photon detection efficiency for each station provide an effective collecting area of 1~m$^2$, and that the detectors can resolve coincidences with resolution of $\tau=0.15$~nsec, both reasonable values practically achievable with contemporary photon detectors \cite{Nomerotski2019, Lee2018, Perenzoni2016}.

Equation~\eqref{eq:vis_factors} confirms the intuitive result, that the two photons will evidence the interference effects we are describing if the time difference $\tau$ between detections and the bandwidth $\Delta\omega$ allowed into the channels satisfy $\tau \, \Delta\omega \le \sim 1$.  We emphasize that this is only an approximate relation as written, and an exact quantitative analysis will depend on the details of the bandwidth and timing resolution profiles.
Schematically, then, we picture the instrument recording photon arrivals at each detector in time bins of width $\tau$, and two hits in the same bin constitute a coincidence.  We can then set the corresponding bandwidth for full interference at $\Delta\nu=\Delta\omega/2\pi=1/2\pi\tau \simeq 1$~GHz.  

We can now describe the basic observational data stream simply as a long series of 0.15~nsec time bins, and if any pair of detectors in the $L$ and $R$ stations each register a hit in the same bin -- accounting for path-length differences as per Equation~\eqref{eq:vis_factors} -- then a pair of that type is counted.  To then estimate the overall pair rate as per Equation~\eqref{eq:flux_counting} we will take the spectral flux density of a magnitude 2 star at wavelength $\lambda = 1$~$\mu$m to be $S=$2000~Jy $\times 100^{-2/5}\simeq 300$~Jy.  This will correspond to a single photon rate in each telescope, from both sources combined, of 
\begin{equation*}
2 \, \frac{300 \, \mathrm{Jy} \,\, 1 \,\mathrm{m}^{2} \,\, 1 \, \mathrm{GHz} }{h c /(10^{-6}\, \mathrm{m})} \simeq 3 \cdot 10^{4}/\mathrm{sec}
\end{equation*}
So the pair occupancy in each 0.15~nsec bin will be quite low, on the order of $10^{-5}$, for an overall pair rate on the order of $\nbar=0.1$~Hz.

With all the values assumed above for the experimental parameters, Equation~\eqref{eq:stdv_deltatheta} yields a resolution on the opening angle of $\sigma[\Delta\theta]\simeq 2 $~mas from one night's observation.  It is interesting to note that the Fisher matrix derivation in Appendix~\ref{app:sensitivity_estimate} holds even in the low-rate case, where the pair rate is significantly slower than the fringe passing rate.

Of course, using only the photons in a single 1~GHz-wide band is a tiny fraction of the information available in the photon field.  As suggested in Figure~\ref{fig:interferometer_spread_detectors_wavefront} we can imagine spectrographically dividing the light from the same objects and carrying out the same measurement in many sub-bands simultaneously.  The full range between, say, $\lambda=0.5\mu$m and $\lambda=1\mu$m spans a bandwidth of $3 \cdot 10^{5}$~GHz.   This plenty of room to deploy, say, $4 \cdot 10^4$ sets of detectors each on its own GHz-wide sub-band, increasing the total rate of observed singles and usable interferometric pairs by this same factor.  

Thinking of each wavelength sub-band as a separate experiment with its own fringe rate, but with all the fringe rates following the scaling in Equation~\eqref{eq:nu_fringe_full}, we can combine the information from the sub-bands statistically.  This will improve the precision on $\Delta\theta$ by the square root of the number of detectors, so deploying $4 \cdot 10^4$ sets of detectors will yield a precision on the order of $\sigma [ \Delta\theta ]\sim10\mu$as for one night's observation in our bright star example.

%
\subsection{Aperture scaling}
\label{subsec:aperture_scaling}
\noindent

With this encouraging result for two very bright objects it is reasonable to highlight the scaling behaviors that could come into play for surveying dimmer objects.  The principal effect of lower object brightness will, of course, be to reduce the number of detected pairs.  As per Equation~\eqref{eq:flux_counting} a reduction in brightness by a factor $\varepsilon < 1$ will result in a factor $\varepsilon^2$ reduction in detected pairs; and as per the  $1/\sqrt{\bar{n}}$ dependence in Equation~\eqref{eq:stdv_deltatheta} this will degrade the precision on the opening angle $\sigma[\Delta\theta]$ by a factor of $1/\varepsilon$.  So, for example, observing a pair of objects which are each dimmer by five visual magnitudes, and with no other changes in the setup, will reduce the brightness by $\varepsilon = 1/100$ and so increase the uncertainty $\sigma[\Delta\theta]$ by $\times$100.  Scaling from the same notional device as in the bright star example above, this implies a precision of $\sim 1$~mas could be reached for a one-night observation of two magnitude~7 objects.

For individual observing stations we can see from Equation~\eqref{eq:flux_counting} that increases in collecting area by some factor $X$ will increase the pair count by $X^2$, and so decrease $\sigma[\Delta\theta]$ by $1/X$.  
We can also see from Equation~\eqref{eq:flux_counting} that
improvements (e.g. reductions) in detector time resolution by a factor of $1/X$ will enlarge each bandwidth $\Delta\nu$ by $X$, while also restricting the coincidence timing, for a net improvement of $X$ in statistics and $1/\sqrt{X}$ in $\sigma[\Delta\theta]$.  So, for example, an improvement in resolution from the 150~psec to 15~psec will decrease $\sigma[\Delta\theta]$ by a factor of $\times1/\sqrt{10}$, or equivalently allow the same precision for sources dimmer by 1.25~visual magnitudes.  

An interesting scaling arises when the same objects are being observed in two cases, but in one case they are dimmer simply by being farther away.  A factor of $X$ increase in distance will, with no other changes, decrease the count of pairs by $1/X^{4}$ and so increase the uncertainty  $\sigma[\Delta\theta]$ by $X^{2}$.  However, as described in Sections~\ref{subsec:extended_sources} and~\ref{subsec:bright_star_example} above the optimal baseline can be chosen as that which will just start to resolve the sources and so go inversely with their size.  This means that in scaling to the same objects at a greater distance away we can increase the baseline by the same factor $X$ without changing the visibility, and thus regain a factor of $1/X$ on $\sigma[\Delta\theta]$ as per Equation~\eqref{eq:stdv_deltatheta}.

Most intriguing, perhaps, is the possibility of increasing sensitivity by deploying a large number of receiving stations.  This would take direct advantage of the fact that no optical connection is needed between stations, keeping the marginal complexity of adding a station quite low.  An array with $N_{\mathrm{St}}$ stations provides $N_{\mathrm{St}}(N_{\mathrm{St}}-1)/2$ pair-wise combinations between them, effectively running this number of experiments in parallel, each with its own baseline.  So, an increase from two stations to, say, 140, would capture about $10^4$ times as many pairs, which could improve $\sigma[\Delta\theta]$ by a factor of $\sim\times$1/100 after proper data collection and processing.  Also, observing across multiple stations brings up the possibility of correcting for atmospheric effects through phase closure methods~\cite{Gottesman2012}, though that is beyond the scope of this paper.

Lastly, we note that it is not the intent of these sections to describe the results obtainable from a real scientific instrument, where any number of systematic effects will come into play.  As a leading example we are not here addressing the effects that atmospheric fluctuations would have on a ground-based experiment.  This is an interesting topic and a full discussion is beyond the scope of this paper.  However, we can note that adaptive optics have been used successfully in compensating for atmospheric effects in astrometric measurements; \cite{Cameron2008, Meyer2011}, and in general they are not fundamental in the same sense as the limitations from available photon statistics.  Also, for very close pairs, where the the objects are being observed in the same aperture and through (nearly) the same isoplanatic patch, the two paths will experience the same atmospheric phase delay, cancelling in the subtraction $\delta_{1}-\delta_{2}$; this is an advantage also enjoyed by HBT measurements.  

\medskip

%
\section{Conclusions}
\label{sec:conclusions}

We have proposed a new type of two-photon interferometry, in which photons from two separate sources are quantum-mechanically interfered at two independent stations. At each station we employ either two independent telescopes or rely on two independent positions on the focal plane of a single telescope.
The basic observables are patterns of correlations between photon detections at the two stations and the overall pattern provides a sensitive measure of the opening angle, e.g. the relative astrometry, of the two sources on the sky.

The scheme is in many ways similar to the intensity interferometry pioneered by Hanbury Brown and Twiss (HBT), but is more general and recovers more information.  In contrast to two-photon intensity interferometry we term the new approach two-photon amplitude interferometry since the photon detections can be both correlated and anti-correlated between the station's detectors.  In a related point the new observable is not specific to thermal sources, being sensitive to the phase differences between two single photons.  An advantage of this new approach, which is also a feature of HBT measurements, is that the two receiving stations do not need to be connected by live optical links but require only slow classical communication channels to compile the correlation observables.  This opens up more flexibility for longer interferometric baselines and thus the prospect of greatly increased precision in astrometry measurements.

We then describe an observational approach in which correlation observables evolve sinusoidally as the time-delay is modulated by the Earth's rotation.  Unlike the case of interferometry for imaging, which requires measuring the amplitude and phase of passing fringes, we show that the astrometric opening angle is sensitive instead to the rate at which the fringes pass, which can be measured with great accuracy.  A basic estimate suggests that a precision on the opening angle on the order of $10\,\mu$as could be achieved in a single night's observation of two bright stars, considering only the irreducible limit from finite photon statistics and ignoring effects of the atmosphere for now.

We also consider the question of how quantum optics techniques can be applied to allow for more flexibility and use of longer baselines for interferometry more generally.  As an example, the distribution of entangled photon pairs to different stations, which can be made robust through the use of polarization or time-bin entanglement, could be easier for long baselines that providing a controlled phase reference.
Once achieved, such distribution would pave the way for using long-distance entanglement as a resource for combining quantum sensor measurements at independent locations more generally. 
We also note that this approach would allow in the longer run to take advantage of the use of quantum memories, as was originally suggested in~\cite{Gottesman2012} and since elaborated by others~\cite{harvard1, harvard2}.  Thus improvements in astrometry may be only the beginning of what can be achieved in astronomical interferometry with the advent of quantum information technologies.


\section*{acknowledgments}
We thank Thomas Jennewein, Duncan England, Emil Khabiboulline, Ning Bao and Eden Figueroa for inspiring discussions.
P.S. and A.N. acknowledge support of the U.S. Department of Energy QuantISED award and BNL LDRD grant 19-30.

%
%
\appendix

%
\section{Classical theory of interferometry}
\label{app:classical}

All three types of interfeometry discussed in this paper, the single photon amplitude interferometry, the single photon intensity interferometry and the two photon interferometry could be, in principle realized in the classical settings with e.g. using radio interferometers. With the advent of modern radio astronomy techniques, there is no real advantage in doing that. However, it is instructive to go through a classical  theory of interferometric imaging, both as a primer to a more complex quantum case, but also to better understand advantages and limitations of various approaches.

The main insight is that classical radiation field from the thermal source can be described as a Gaussian random field, which is stationary in time (i.e. correlators are diagonal in the frequency domain) and statistically independent between various directions on the sky. The traditional amplitude interferferometry depends on measuring the two-point statistics of this field. The intensity interferometer of the HBT-type measures the 4-point correlator from the same direction, which results in the non-zero mean signal, but Gaussian fluctuations around this mean. The two-source intensity interferometry relies on four-point function combining the intensity from two direction in the sky: this signal has again a zero mean but fluctuations that encode the quantities of interest.

We will work with a thermal source observed around a frequency $\omega$ with a bandwidth of $\Delta \omega$, where we work in a narrow bandwidth limit $\Delta \omega \ll \omega$. This immediately introduces two time-scales into the problem, a single wavelength timescale $\omega^{-1}$ and a coherence timescale $\Delta \omega^{-1}$. We will model the single polarization of electric field as a complex correlated Gaussian field, where the complex components encode the phase of the field with respect to an arbitrary time reference\footnote{This could be viewed, for example, as simple components of an appropriate Fourier transform or polyphase filterbank samples. The calculation can be self-consistently performed also using the real-valued fields (e.g. the component of the electric field $E_x(t)$), which changes some of the correlators, but gives the same final results.}. The electric field coming from sky position $\theta$ at time $t$ has the correlators:
\begin{eqnarray}
  \left < E(\theta, t) \right>  &=& 0  \\
  \left < E(\theta, t) E^*(\theta', t') \right>  &=& I(\theta) \delta^D(\Delta \theta) e^{i\omega \Delta t} g \left (\Delta \omega \Delta t\right)  \label{eq:class}
\end{eqnarray}
where $g$ is the square of the Fourier transform of the bandpass function with $g(0)=1$, $g(\omega^{-1}\Delta \omega )\sim 1$ and $g(x)\rightarrow 0$ for $x\rightarrow \infty$. The higher order correlators can be derived using Wick theorem. This is, in short, stating that emission from different points on the sky is independent of each other and that emission from the same point on the sky has the expected correlations for a Gaussian field with a known power spectrum.  
\newcommand{\bb}{\mathbf{B}}
\newcommand{\bu}{\mathbf{u}}

We imagine the same basic setup as depicted in the Figure \ref{fig:interferometer_spread_detectors_wavefront}. Two stations labelled $L$ and $R$ are observing two nearby sources 1 and 2, but we replace both beam splitters with 4 analog detectors $a$, $b$, $e$ and $f$.  We assume that the signals from the two sources can be separated optically, i.e. using several feeds in case of radio astronomy or two optical fibers. The instrument response for a signal from each source is described by the corresponding beams. For simplicity we assume beams to be normalized to the effective solid angle and have no overlap, i.e.
\begin{eqnarray}
  \int |B^2_1(\theta)| d^2\theta = \int |B_2(\theta)|^2 d^2\theta &=& \Omega\\
  \int X(\theta) B_1(\theta) B_2^*(\theta)  d^2\theta &=& 0
\end{eqnarray}
for any function $X(\theta)$.  Note that in general beams can be complex.

The electric fields for all four measurements is then given by
\begin{eqnarray}
  E_{a}  &=& \int E(\theta, t) B_1(\theta) d^2\theta \\
  E_{b}  &=& \int E(\theta, t) B_2(\theta) d^2\theta \\
  E_{e}  &=& \int E(\theta, t+ c^{-1} \bb \cdot \theta) B_1(\theta) d^2\theta\\
  E_{f}  &=& \int E(\theta, t+ c^{-1} \bb \cdot \theta) B_2(\theta) d^2\theta
\end{eqnarray}
where we have arbitrarily set the phase center of the signal to be at the receiver L.

Different approaches to interferometry simply correspond to different correlators of the input field. 

\subsection{Amplitude interferometry}

In this case we correlate just the signal corresponding to the source $A$. Visibility is given by 
\begin{equation}
  V_\bb = \left<E_a E_e^*\right> 
\end{equation}
Plugging in correlation in Equation~\eqref{eq:class}, assuming the path differences are very small compared to the correlation length and simplifying we arrive at the well know result
\begin{equation}
V_{\bb} = \int  I(\theta) e^{-i 2\pi\ \bu \cdot \theta } |B_1(\theta)|^2  d^2\theta = FT[I_A],
\end{equation}
where $\bu=\bb/\lambda$ and $I_A (\theta) = I(\theta) |B_1(\theta)|^2$. In other words, we recover the well known result that the traditional optical amplitude inteferometry measures the Fourier transform the of intensity field on the sky.

\subsection{Intensity (HBT-type) Interferometry}

In intensity correlation, we are interested in intensities  of the signal, that is
\begin{equation}
I_{x} (t) = E_{x} E_{x}^{*},
\end{equation}
where $x$ can be any of a, b, e, f.

The average intensity is given by 
\begin{equation}
I_1 = \left<I_{a} \right> =   \left<I_{e} \right> = \int I(\theta) B_1^2(\theta) d^2\theta 
\end{equation}
The visibility is then given by the four-point function
\begin{equation}
  V_\bb = \left<I_{a} I_{e}\right> - \left<I_{a}\right> \left<I_{e}\right>
\end{equation}
After some manipulation, we arrive at
\begin{eqnarray}
 V_\bb & = & \iint I(\theta) I(\theta') e^{-i 2\pi\ \bu \cdot (\theta-\theta') } |B_1(\theta)|^2 |B_1(\theta')|^2  d^2\theta d^2\theta' = FT[I_A]FT^*[I_A]
\label{eq:HBT_visibility}
\end{eqnarray}
Intensity interferometry or HBT experiment measures the modulus of the Fourier transform of the source plane. HBT can be derived in full from simply classical fields. It relies on measuring the variance of the intensity compared to mean intensity, which in turn encodes the spatial structure of the source. However, the phase information is lost.

\subsection{Two Source Amplitude Interferometry}

In the two source interferometry, we instead take the signal cross-correlation 
\begin{eqnarray}
  X_{ab} &=&  E_{a} E_{b}^{*},\\
  X_{ef} &=&  E_{e} E_{f}^{*}.
\end{eqnarray}
Note that the quantities $I_{ab}$ and $I_{ef}$ are complex, compared
to normal intensities described above which are manifestly real.

Because the sources $1$ and $2$ are independent and beams do not overlap, we have
\begin{equation}
  \left<X_{ab}\right> = \left< X_{ef}\right> = 0
\end{equation}
However, the four point function does not vanish. We find
\begin{eqnarray}
  V_\bb  =  \left<X_{ab} X_{ef}^* \right> = \left<E_{a} E_{b}^{*} E^{*}_{e} E_{f}\right>  = \iint  I(\theta) I(\theta') e^{-i 2\pi \bu \cdot (\theta-\theta')}  d^2\theta d^2\theta'
 =  FT[I_1] FT^*[I_2]
\label{eq:four_point_classical}
\end{eqnarray}
This seems to be a straight-forward generalization of the intensity interferometry discussed in the previous section: the visibility is given by the product of the Fourier transforms of sources $1$ and $2$ (with later being conjugated). However, the important difference is that we can choose sources $1$ and $2$. For example, by choosing source $2$ to be a point source (either real or artificial), the two source interferometry directly measures the Fourier transform of the source $1$ much like amplitude interferometry. This option does not exist in the traditional intensity interferometry. Much of this intuition carries over to the quantum two photon interferometry, however with important differences in the way the signal is operationally measured.

%
\section{Field Propagation}\label{app:field}
To describe the effects of field propagation we will employ methods of Fourier optics and Green function formalism by introducing a function of field propagator. The propagator allows to find the distribution of the field in the observation plane based on the field distribution in the source plane. Commonly, it is assumed that each source consists of many independent point-like sub-sources. Thus, according to the superposition principle we can describe the field propagation by the following "weight" function \cite{goodman2005}:
\begin{eqnarray}\label{g_spatial-a}
G_{\omega}(\vec{r}|\vec{r}_{0,\perp}^{(j)}) \approx\frac{(-i\omega) e^{i\frac{\omega R_{j}(\vec{r}_{0\perp}^{(j)})}{c} }}{2\pi c R_{j}(\vec{r}_{0\perp}^{\ (j)})},
\end{eqnarray}
where $R(\vec{r}_{0\perp}^{j}) = \sqrt{\vec{R}_{j}^2 - 2\left(\vec{R}_{j}\vec{r}_{0\perp}^{j}\right)+{\vec{r}}^{2 \ (j)}_{0\perp}}$ is the distance between the sub-source at $\vec{r}_{0\perp}$ and observation station (L or R). 
In the far-field approximation, one can assume with very good accuracy that the denominator in \eqref{g_spatial-a} is equal to $R_{j} \approx R_{j}(0)$, the characteristic distance between the object and observation planes, and then expand $R$ as:
\begin{eqnarray}\label{distance-a}
R(\vec{r}_{0\perp}) \approx \left(R_{j} - \frac{\left(\vec{R}_{j}\vec{r}_{0\perp}\right)}{R_{j}}+\frac{{\vec{r}_{0\perp}}^{2}}{2R_{j}}\right).
\end{eqnarray}
We note that, in general, we cannot neglect the second term of the above expansion when substituting $R(\vec{r}_{0\perp})$ in \eqref{g_spatial-a} simultaneously accounting for the phase factor $e^{i\frac{\omega R(\vec{r}_{0\perp})}{c} R(\vec{r}_{0\perp})}$. We can further rewrite the positive part of the electric field operator (see Equation~\ref{propagation} in the main text) at the observation plane in the following form:
\begin{eqnarray}\label{field_full-a}
\hat{E}^{[j](+)}(\vec{R}_{js},t) \approx \frac{1}{R_{js}}\int \sqrt{\frac{\hbar\omega^3}{(2\pi)^3c^2}}
\int_{S_{j}} \hat{a}^{[j]}_{\vec{k}_{\perp}}(\omega) e^{i\left(\vec{k}_{\perp} - \frac{\omega}{c}\frac{\vec{R}_{js}}{R_{j}}\right)\vec{r}_{0\perp}^{(j)}}e^{i\frac{\omega\vec{r}^{2 \ (j)}_{0\perp}}{2cR_{j}}}d\vec{r}^{(j)}_{0\perp} e^{i\left(\omega\frac{ R_{js}}{c}- t\right)} d\vec{k}_{\perp}d\omega,
\end{eqnarray}
where we denoted $\vec{R}_{js}$ as a vector connecting each source ($j = 1,2$) and each observation station ($s = L,R$). We also assumed that $|\vec{R}_{js}|\approx R_{j}$ in the denominators of  $\frac{\omega c_{\omega} e^{i\omega R_{js}}}{2 c \pi R_{j}}$ in  \eqref{g_spatial-a} and in \eqref{field_full-a} after substitution of expanded $R$. Using this notation and taking into account all simplifications we can write $\vec{R}_{jR} -\vec{R}_{jL} = \vec{B} $ and $|R_{jL} - R_{jR}| = B \sin{\theta_{j}}$, which is obvious from Figures \ref{fig:single_photon_beamsplitter} and \ref{fig:geometry}.

For improved simplicity we can reorganize \eqref{field_full-a} and introduce a new function:  $\tilde{G}^{[j]}_{\vec{R}_{js}}(\omega,\vec{k}_{\perp})$ as follows:
\begin{eqnarray}\label{simplified-g-a}
\tilde{G}^{[j]}_{\vec{R}_{js}}(\omega,\vec{k}_{\perp}) = \sqrt{\frac{\hbar \omega^3}{(2 \pi)^3 R_{js}^2 c^2}} 
\int_{S_{j}}e^{i\left(\vec{k}_{\perp} - \frac{\omega}{c}\frac{\vec{R}_{js}}{R_{js}}\right)\vec{r}_{0\perp}^{(j)}}e^{i\frac{\omega\vec{r}^{2}_{0\perp}}{2cR_{j}}}d\vec{r}_{0\perp}^{(j)}.
\end{eqnarray}
Using Equation \eqref{field_full-a} one can also describe the superposition of fields from two sources arriving at the beam-splitter (BS) for each observation station $s = L,R$, see Figure \ref{fig:single_photon_beamsplitter}.  One can then describe the output field operators after the BS transformation for each observation station through the below input operators:
\begin{eqnarray}\label{field_operator_init-a}
&&\hat{E}^{(+)}_{a_{s}} = \frac{1}{\sqrt{2}}\left(\hat{E}^{{[1]}{(+)}}_{s}+ (-1)^{a_{s}}\hat{E}^{{[2]}{(+)}}_{s}\right) =  \nonumber \\ 
&&\int \frac{\tilde{G}^{[1]}_{\vec{R}_{1s}}(\omega,\vec{k}_{\perp})\hat{a}^{[1]}(\vec{k}_{\perp},\omega)}{\sqrt{2}}e^{i\omega (\frac{R_{1s}}{c} - t+\delta_{1s})} d\vec{k}_{\perp}d\omega +
(-1)^{a_{s}}\int \frac{\tilde{G}^{[2]}_{\vec{R}_{2s}}(\omega,\vec{k}_{\perp})\hat{a}^{[2]}(\vec{k}_{\perp},\omega)}{\sqrt{2}}e^{i\omega (\frac{R_{2s}}{c} - t+\delta_{2s})} d\vec{k}_{\perp}d\omega, \nonumber \\
&& a_{s}\in \{0,1\}, \ \forall s = {\rm{L,R}}
\end{eqnarray}
where index $ s =\rm{L,R}$ is referred to the observation stations and index ${a}_{s} = 0,1$ parameterizes output ports of both BSs: $({a = 1}_{L},{a =2}_{L}) \equiv  (c,d)$, $ ({a = 1}_{R},{a =2}_{R}) \equiv  (g,h)$. $\delta_{j,s}$ are the additional phase delays before the BSs with each delay corresponding to $\delta_{(1,2)L} = \frac{\Delta L_{(a,e)}}{c}$ and $\delta_{(1,2)R} = \frac{\Delta L_{(b,f)}}{c}$. 
To describe the two-photon correlation we will use the explicit fourth-order correlation function defined in the main text in Equation~\ref{corr.avarage}, also adding to it Equation~\ref{field_operator_init-a}. By performing the time and normal ordering in Equation~\ref{corr.avarage} we can rewrite it in the following form :
\begin{eqnarray}\label{corr.avarage-a}
&&\Gamma_{1,2}^{a_{L},a_{R}}(\tau) = \braket{\hat{E}^{{(-)}}_{a_{L}}(t)\hat{E}^{{(-)}}_{a_{R}}(t + \tau)\hat{E}^{{(+)}}_{a_{R}}(t+\tau)\hat{E}^{{(+)}}_{a_{L}}(t)} = \nonumber \\
&& \braket{\hat{E}^{{[1]}{(-)}}_{L}\hat{E}^{{[1]}{(-)}}_{R}\hat{E}^{{[1]}{(+)}}_{R}\hat{E}^{{[1]}{(+)}}_{L}}+ \braket{1\Longleftrightarrow 2} + \braket{\hat{E}^{{[1]}{(-)}}_{L}\hat{E}^{{[2]}{(-)}}_{R}\hat{E}^{{[2]}{(+)}}_{R}\hat{E}^{{[1]}{(+)}}_{L}} + \braket{L\Longleftrightarrow R} + \nonumber \\
&&+\left(-1\right)^{a_{L}+a_{R}}\bigg[\braket{\hat{E}^{{[1]}{(-)}}_{R}\hat{E}^{{[2]}{(-)}}_{L}\hat{E}^{{[2]}{(+)}}_{R}\hat{E}^{{[1]}{(+)}}_{L}} + {\rm{c.c.}}\bigg].
\end{eqnarray}
Symbols $\braket{1\Longleftrightarrow 2}$ and $\braket{L\Longleftrightarrow R}$ indicate calculation of the same expression as on the left-hand side where the corresponding indices $1,2$ and $L,R$ are interchanged, the symbol $\rm{c.c.}$ stands for the complex conjugate operation. 
Deriving \eqref{corr.avarage-a} we implied that unpaired terms like $\hat{E}^{{[1]}{(-)}}\hat{E}^{{[1]}{(-)}}\hat{E}^{{[2]}{(+)}}\hat{E}^{{[1]}{(+)}}$ or $\hat{E}^{{[1]}{(-)}}\hat{E}^{{[2]}{(-)}}\hat{E}^{{[2]}{(+)}}\hat{E}^{{[2]}{(+)}}$ are equal to zero \cite{ou2007}, which is a common property of thermal radiation \cite{mandel_wolf1995}. By substituting \eqref{field_operator_init-a} in \eqref{corr.avarage-a}, and employing simplifications mentioned in deriving Equations~\ref{field_full-a} and \ref{simplified-g-a}, we obtain Equation~\eqref{corr.rewritten} from the main text as our final result.

%
\section{Sensitivity Estimation}
\label{app:sensitivity_estimate}

Let us consider pair counts with rate given by $n(t)$. We further assume $n(t)$ is a function of some theory parameters $\theta_i$.   We will take the Poisson limit: we will start by taking finite but small  time resolution $\Delta t$ window in which we either have an event with probability $n(t)\Delta t \ll 1$ or we do not have one with a probability $1-n(t)\Delta t$. The total likelihood is the product over these probabilities
\begin{equation}
  \mathcal{L} = \prod_{t_i \in \mbox{events}} n(t_i) \Delta t \prod_{t_i \notin \mbox{events}} (1-n(t_i)\Delta t),
\end{equation}
where the first product is over all time stamps where $t_i$ has an event and the second where it has not.
To proceed with the Fisher prescription, we take log likelihood, which converts the product into a sum and take a derivative twice to obtain
\begin{equation}
  \frac{\partial^2 \log \mathcal{L}}{\partial \theta_i \partial \theta_j} =
  \sum_{t_i \in \mbox{events}} \left[ \frac{1}{n^2}\frac{\partial{n}}{\partial \theta_i} \frac{\partial{n}}{\partial \theta_j} + \frac{1}{n} \frac{\partial^2 n}{\partial \theta_i \theta_i}  \right]_{t_i}
+ \sum_{t_i \notin \mbox{events}} \left[ \frac{\Delta^2 t}{(1-n\Delta t)^2}\frac{\partial{n}}{\partial \theta_i} \frac{\partial{n}}{\partial \theta_j} - \frac{\Delta t}{(1-n\Delta t)} \frac{\partial^2 n}{\partial \theta_i \theta_i}  \right]_{t_i}
\end{equation}
Now we do the final manipulation. We first take the average over realizations. This means that event in the time-slot $t_i$ will occur precisely with its fiducial rates, which allows us to say
\begin{eqnarray}
  \left< \sum_{t_i in \mbox{events}} \left(\cdots\right) \right> &\rightarrow& \sum_{t_i} n(t_i) \Delta t \left(\cdots\right), \\
  \left< \sum_{t_i \notin \mbox{events}} \left(\cdots\right) \right>  &\rightarrow& \sum_{t_i} (1-n(t_i) \Delta t) \left(\cdots\right). 
\end{eqnarray}
Next we take continuum limit by keeping the first order in $\Delta t$ and converting
\begin{equation}
\sum_{t_i} \Delta t \rightarrow \int dt.
\end{equation}

We thus arrive at the Fisher matrix for a Poisson process as
\begin{equation}
  F_{ij} = \int dt \frac{1}{n(t)} \frac{\partial n(t)}{\partial \theta_i} \frac{\partial n(t)}{\partial \theta_j}
\end{equation}

Note that strictly speaking this is an approximation to the problem we are trying to solve. When looking for a coincidence we look for pairs in gated time windows. In each $\Delta t$ there is either one event or zero, but in any Poisson process there is at least a notional possibility of having two or more events in an arbitrarily small window. A proper calculation then becomes analytically untractable, but can be approximated in series expanded in the small parameter $n\Delta t$. The Poisson result is then a leading term in that series.

In our case, the signal is given by Equations \eqref{eq:f1} and \eqref{eq:f2} with variables of interest $\nbar$, $V$, $\omega_f$ and $\phi_f$, but we are really just interested in $\omega_f$.

The rest involves turning the crank and doing the algebra. The Fisher matrix integral involving a power law in $t$ and a periodic function $P(\omega_f t+\phi)$. We assume that the fringing is fast, which allows  these integrals to be approximated assuming $t$ does not change significantly over a single oscillation:
\begin{equation}
  \int t^\alpha P(\omega_f t + \phi) dt \approx \int t^\alpha dt \left<P(\omega_f t' +\phi) \right>_{t'}
\end{equation}

With this approximation we find $F_{\omega_f \nbar}=0$, $F_{\omega_F V}=0$, $F_{\phi_f \nbar}=0$ and $F_{\phi_f V}=0$. Note that this does not mean that values of those parameters do not impact the error on $\omega_f$, but that measurements of these quantities is uncorrelated with the measurement of $\omega_f$. We have three remaining relevant quantities:
\begin{eqnarray}
  F_{\omega_f \omega_f} &=& \frac{1}{6} \nbar V^2 \, T^3 \, \kappa(V),\\
  F_{\omega_f \phi_f} &=& \frac{1}{4} \nbar V^2 \, T^2 \, \kappa(V),\\
  F_{\phi_f \phi_f} &=& \frac{1}{2} \nbar V^2 \, T \, \kappa(V)
\end{eqnarray}
where $\kappa(V)$ is a small auxiliary function with value order one
\begin{equation}
    \kappa(V) = \frac{1}{2\pi} \int_0^{2\pi} \frac{\cos^2x -1}{V^2 \cos^2 x -1} dx = \frac{1-\sqrt{1-V^2}}{V^2}
    \label{eq:aux_func_S}
\end{equation}
\noindent
bounded within $\kappa(0)=\frac{1}{2}$ and $\kappa(1)=1$.

The marginalized error squared, ie the variance, on $\omega_f $ is thus
\begin{equation}
  \sigma^2 \left[ \omega_f  \right] = \frac{24}{\nbar T^3 V^{2} \kappa(V)}.
  \label{eq:variance_omega_f}
\end{equation}

%


\bibliography{QApaper_OJA_Final.bib}{}
\bibliographystyle{aasjournal}

\end{document}